\documentclass[3p,  number, single column]{elsarticle}

\usepackage{graphicx}
\usepackage{epsfig}
\usepackage{cite}
\usepackage{amsmath}
\usepackage{amssymb}
\usepackage{url}
\usepackage{fancyvrb}
\usepackage{setspace}

\newcommand{\be}{\begin{equation}}
\newcommand{\ee}{\end{equation}}
\newcommand{\bea}{\begin{eqnarray}}
\newcommand{\eea}{\end{eqnarray}} 
\newcommand{\nn}{\nonumber}
\newcommand{\cs}{\mathbb S}
\newcommand{\GeV}{{\rm\ GeV}}

\newcommand{\TeV}{{\rm\ TeV}}

\begin{document} 

\begin{frontmatter}

\title{MadDM v.1.0: Computation of Dark Matter Relic Abundance Using MadGraph5}

\author{Mihailo Backovi\'{c}}
\ead{mihailo.backovic@weizmann.ac.il}
\address{Department of Particle Physics and Astrophysics, Weizmann Institute of Science, Rehovot 76100, Israel }

\author{Kyoungchul Kong}
 \ead{kckong@ku.edu}
\address{  Department of Physics and Astronomy,  University of Kansas, Lawrence, KS 66045, USA}

\author{Mathew McCaskey}
\ead{mccaskey@ku.edu}
\address{  Department of Physics and Astronomy,  University of Kansas, Lawrence, KS 66045, USA}



\begin{abstract}
We present MadDM v.1.0, a numerical tool to compute dark matter relic abundance in a generic model. The code is based on the existing MadGraph 5 architecture and as such is easily integrable into any MadGraph collider study. A simple Python interface offers a level of user-friendliness characteristic of MadGraph 5 without sacrificing functionality. MadDM is able to calculate the dark matter relic abundance in models which include a multi-component dark sector, resonance annihilation channels and co-annihilations. We validate the code in a wide range of dark matter models by comparing the relic density results from MadDM  to the existing tools and literature.  
\end{abstract}


\end{frontmatter}

\begin{center}
\includegraphics[scale = 0.6]{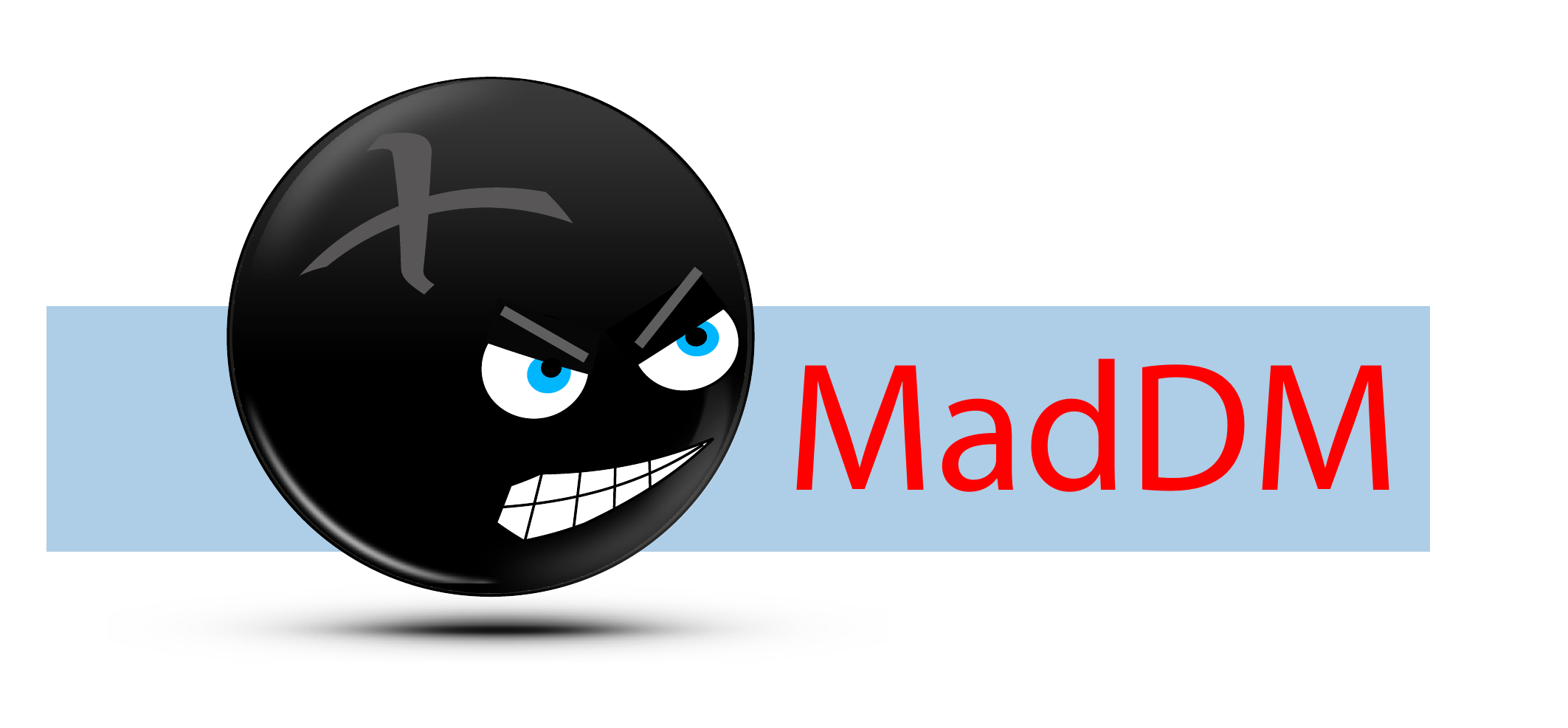}
\end{center}

\newpage
\section{Introduction}
\label{sec:intro}

\large
\onehalfspacing
A high demand for computational tools in collider physics resulted in a myriad of monte-carlo event generators. Much care and effort has also been dedicated to link mass spectrum generators, detector simulators, and parton showering models with the existing tools. On the other hand, the experimental effort of dark matter (DM) searches sparked a demand for another type of numerical tool aimed at dark matter phenomenology. The gap between dark matter tools and collider tools has not been completely bridged, with very few numerical packages offering the ability to perform both collider and dark matter calculations. 

Perhaps the only set of tools able to integrate dark matter phenomenology with collider physics is  \verb|CalcHEP| \citep{Belyaev:2012qa} and \verb|micrOMEGAs| \citep{Belanger:2006is}, while other popular collider tools such as \verb|MadGraph| \citep{Alwall:2011uj} and \verb|Sherpa| \citep{Gleisberg:2008ta} lack this ability. In addition, model specific dark matter tools such as \verb|DarkSusy| \citep{Gondolo:2004sc} allow for detailed calculations of galactic dark matter relic signals in the framework of super-symmetric models but without the ability to \textit{easily} integrate into collider tools. 

\verb|MadDM| aims to bridge the gap between collider oriented event generators and dark matter physics tools. We chose to build \verb|MadDM| on the existing \verb|MadGraph 5| infrastructure for two reasons. First, \verb|MadGraph| is widely used among the experimental collaborations in searches for Beyond the Standard Model (BSM) physics and we wish to provide them with the ability to easily incorporate relic density constraints into searches for models which support a dark matter candidate. Second, the Python implementation of the \verb|MadGraph 5| code allows for a fairly straightforward development of \verb|MadGraph| add-ons without sacrificing functionality.  

Very much like \verb|MadGraph/MadEvent|, the \verb|MadDM| code is split into the process generating module written in Python and a FORTRAN numerical module for calculating relic abundance. The Python module automatically determines the dark matter candidates from a user specified model and generates relevant dark matter annihilation diagrams. The numerical FORTRAN code then solves the dark matter density evolution equation for a given parameter set and outputs the resulting relic density. We provide a simple, \verb|MadGraph 5| inspired interface for the convenience of the user, as well as the ability to perform multi-dimensional parameter scans.

\verb|MadDM| is compatible with any existing \verb|MadGraph 5| model, as well as model-specific mass spectrum or decay width calculator which can produce a Les Houches formatted MadGraph parameter card. The code is able to handle the full effects of co-annihilations and resonant annihilation in a generic model. In models which contain more than one dark matter particle, \verb|MadDM| can solve either the full set of coupled density evolution equations or a single equation involving effective thermally averaged cross sections. The former gives \verb|MadDM| the ability to calculate relic density in models which allow for dark matter conversions \citep{Aoki:2012ub}.

We use \verb|micrOMEGAs| as a benchmark for \verb|MadDM| validation. Standard Model extensions with singlet scalar fields serve as simple tests of the \verb|MadDM| performance in models which feature resonant annihilation channels and co-annihilations. In addition, we also consider more complex models such as the Minimal Supersymmetric Standard Model (MSSM) and Minimal Universal Extra Dimensions (MUED). We find good agreement between \verb|MadDM| and \verb|micrOMEGAS| in a wide range of dark matter models, with the exception of regions of the MSSM parameter space where details of the running $b$ mass can cause large effects.

We begin in Section 2 by reviewing relic abundance calculation and introduce \verb|MadDM| in Section 3.
We reserve Section 4 for validation and include an example for MadDM Python script in Appendix.

\section{Calculation of Relic Abundance}

\subsection{The Density Evolution Equation with Co-annihilations}

We begin with a simple case in which a single dark matter particle is stable and annihilates only with itself~\footnote{The case where dark matter has a unique anti-particle is a straightforward extension.}, while we postpone the discussion of a more general treatment of dark matter relic density until the following sections.  The calculation of relic abundance for a dark matter candidate $\chi$ is governed by the density evolution equation
\begin{equation}
\frac{dn_{\chi}}{dt}+3Hn_{\chi} = -2\langle\sigma_{eff}v\rangle\left(n_{\chi}^{2}-(n_{\chi}^{EQ})^2\right),
\label{eq:density_evolution0}
\end{equation}

\noindent where $n_X$ is the number density of dark matter particles, and  $H$ is the Hubble expansion constant. The indexes $i, j$ run over all the co-annihilating particles partners in the model, while $\langle\sigma_{eff}|v|\rangle$ is the effective thermally averaged annihilation cross section
~\footnote{The factor of $2$ represents the fact that there are either two dark matter particles being created or destroyed in the annihilation/creation process. This factor is usually canceled with an additional factor of $1/2$ which comes from double counting in the initial state phase space.  Our definition of the thermally averaged annihilation cross section will explicitly have this factor for clarity.}
\begin{eqnarray}
\langle\sigma_{\text{eff}}v\rangle &\equiv& \sum_{i,j=1}^{N}\langle\sigma(\chi_{i}\chi_{j}\to SM)v\rangle\frac{n_{\chi_{i}}^{EQ}n_{\chi_{j}}^{EQ}}{(n_{\chi}^{EQ})^{2}},\nonumber \\
\langle\sigma_{\chi\bar{\chi}\leftrightarrow X\bar{X}}|v|\rangle &=& \frac{2Tm_{\chi}^{2}}{(2\pi)^{4}n_{eq}^{2}(1+\delta_{\chi\bar{\chi}})}\int_{0}^{1} d\beta~\frac{\beta}{(1-\beta^{2})^{2}} \nn \\
 &\times& \sqrt{\frac{\lambda(s,m_{\chi}^{2},m_{\bar{\chi}}^{2})}{s}}~K_{1}\left(\frac{\sqrt{s}}{T}\right)W_{ij}(s), \label{eq:sigv}
\end{eqnarray}
where $k$ runs over all the particles in the process, while $n_{\chi}^{EQ}$ is the thermal equilibrium density of the dark matter species with mass $m_{\chi}$ and $g_{\chi}$ internal degrees of freedom at temperature $T$:
\begin{equation} 
\label{eq:Neq}
n_{\chi}^{EQ}\approx g_{\chi}\left(\frac{m_{\chi}T}{2\pi}\right)^{3/2}e^{-m_{\chi}/T}.
\end{equation}

$K_1$ is the modified Bessel function of the second kind, $\beta \equiv \sqrt{1 - 4m_\chi^2 / s}$ is the relativistic velocity of the initial state particles, while the numerically convenient $W_{ij}$ proxy is defined as:
\begin{eqnarray} \label{eq:wij}
W_{ij}(s)&=& \frac{\sqrt{\lambda(s,m_{X}^{2},m_{\bar{X}}^{2})}}{(1+\delta_{X\bar{X}})8\pi s}\int\sum_{\rm spins}|\mathcal{M}|^{2}\left(\frac{d\Omega_{CM}}{4\pi}\right),
\end{eqnarray}

\noindent where $\lambda(a,b,c)$ is the usual two body kinematic function:
\be
\lambda(a,b,c) \equiv  a^2+b^2+c^2-2(ab+bc+ac).
\ee

Notice that the definition of effective thermally averaged cross section reduces to~$\langle \sigma(\chi \chi \to SM) |v| \rangle$ in case of only one dark matter particle and no-coannihilating partners. 

To calculate the dark matter relic abundance one must solve Eq. \ref{eq:density_evolution0}. Due to the presence of both  $n$ and $n^2$ terms, the numeric solution to Eq. \ref{eq:density_evolution0} in the present form is not practical. Tracking the number density of dark matter particles per unit entropy $(Y \equiv n_{\chi}/s)$ instead of $n_{\chi}$ \footnote{The change of variables assumes that the universe undergoes an adiabatic expansion}, as well as substituting temperature for $x\equiv m_\chi / T$ allows for Eq. \ref{eq:density_evolution0} to be cast into a simpler form which \verb|MadDM| uses to solve for relic density:

\begin{eqnarray}
\label{eq:density_evolution_numerical} 
\frac{dY}{dx}&=&-\sqrt{\frac{\pi}{45~g_{*}}}\frac{g_{*S}~m_{pl}~m_{\chi}}{x^{2}}\langle\sigma_{eff}|v|\rangle[Y^{2}-Y^{2}_{EQ}] , \\
Y_{EQ}(x)&\approx&\frac{45}{4\pi^{4}}\sqrt\frac{\pi}{2}\frac{g}{g_{*S}}x^{3/2}e^{-x} \qquad (x \gg 1).
\end{eqnarray}

It is possible to find approximate numerical solutions to Eq. \ref{eq:density_evolution_numerical} by integrating the equation from the freeze-out temperature $x_f$ to infinity, whereas a more precise solution can be obtained by fully integrating Eq.~\ref{eq:density_evolution_numerical} (as explained in Section~\ref{sec:boltzprac}). The result of the calculation is  $Y_{\infty}$, the number of dark matter particles per unit entropy in the present universe. 
$Y_\infty$ is  related to the relic density parameter $\Omega h^2$ as
\begin{equation}
\Omega h^{2}\equiv \frac{\rho_\chi} {\rho_c} = \frac{m_{\chi}~Y_{\infty}~s_{\infty}}{1.05\times 10^{-5}~\text{GeV}^{2}~\text{cm}^{-3}},
\end{equation}
where $\rho_c$ is the critical energy density of the universe, $\rho_\chi$ is the energy density of dark matter and $s_\infty \approx  2889.2\, {\rm cm}^{-3}$ is the entropy density in the present universe.

\subsection{Thermal Freeze-out and the Numerical Integration of the Density Rate Equation}
\label{sec:boltzprac}

The density evolution equation should, in principle,  be integrated from the beginning of the Universe (appropriately at $t=0$) to today ($t=\infty$).  In a numerical sense this is not possible, much less practical. As we will show momentarily, the differential equation only needs to be integrated over a much smaller range.  

The thermal evolution of the dark matter in the early universe, $x \sim O(1),$ is characterized by a rapid dark matter annihilation rate
\be
	\Gamma_\chi \equiv \langle \sigma_A|v|\rangle \, n^{EQ}_\chi.
\ee
The expansion rate of the universe (characterized by the Hubble parameter $H$) in the low $x$ regime is much lower than $\Gamma_\chi$ due to the large suppression of $1/m_{pl}$. 
In addition, the dark matter annihilation rate at high temperature is compensated by the inverse process (creation rate),  thus guaranteeing that dark matter remains in thermal equilibrium; \textit{i.e.} $Y \approx Y_{EQ}$.

As the universe cools down and $x$ increases, the annihilation rate $\Gamma_\chi$ obtains an exponential suppression. At some point in thermal evolution, $H > \Gamma_\chi,$ meaning that the annihilation rate effectively stops and the number density of dark matter becomes constant. The so-called  ``thermal freeze-out'' occurs roughly at
\be
		H \approx \Gamma_\chi,
\ee
with the resulting temperature called the ``freeze-out temperature'' \footnote{Note  that the in cases in which $\langle \sigma_A|v|\rangle$ has a non-trivial dependence on $x$, the temperature at which dark matter density $Y$ decouples from the equilibrium distribution and the temperature at which the relic density freezes out can be different.  Resonant dark matter annihilation is a notable example, where large resonance widths can prolong the efficient dark matter annihilation rates \citep{Griest:1990kh,Ibe:2008ye,Guo:2009aj}.}.

\begin{figure}[tbp]
\begin{center}
\includegraphics*[angle=0,width=0.7\textwidth]{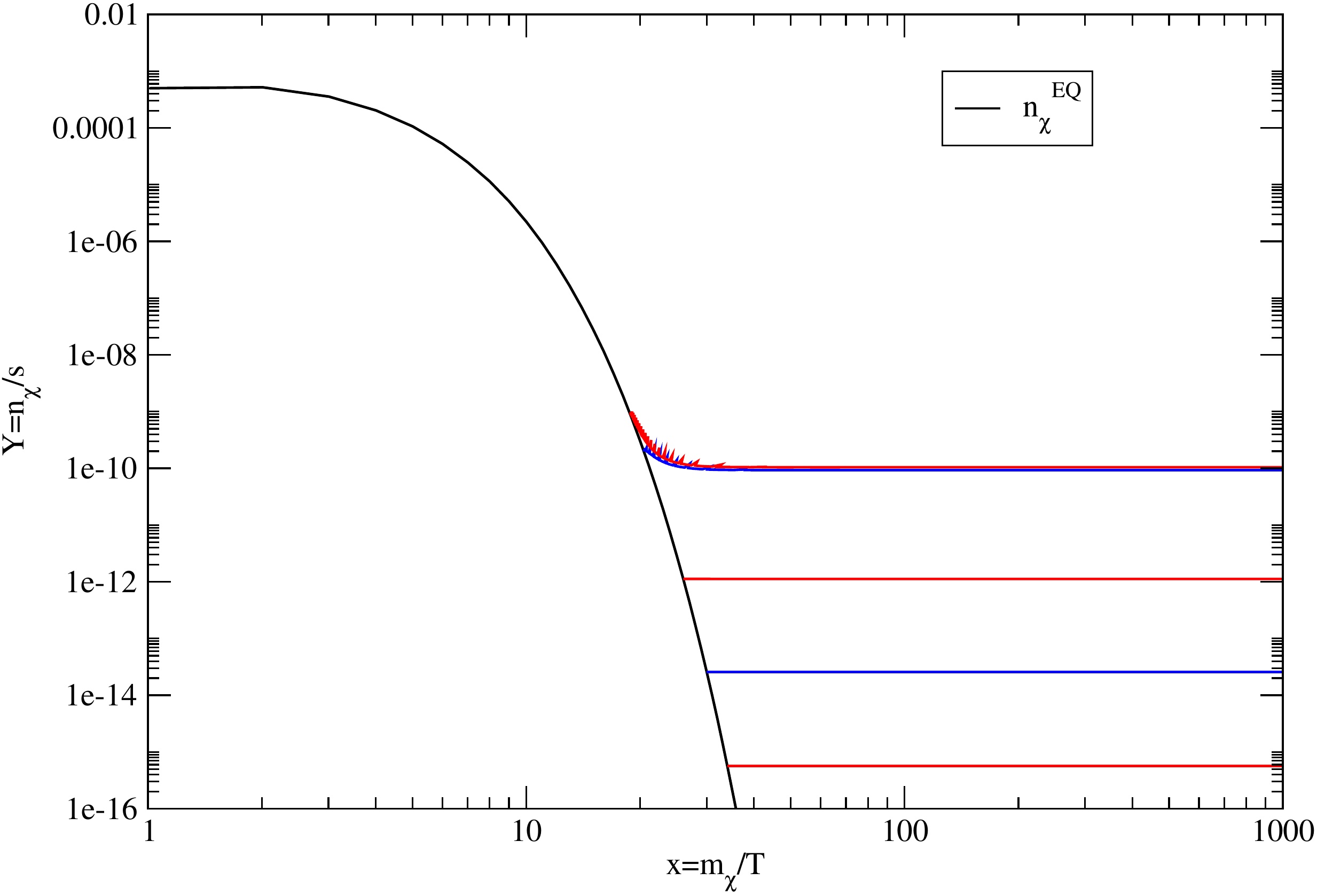}
\caption{Practical integration of the density evolution equation without knowing the freeze-out temperature before hand.  The alternating red and blue lines show the solution starting at smaller initial values for $x=m_{\chi}/T$.  Once two consecutive solutions are within a specified percent error the solution is considered the correct one.}
\label{fig:ode_int}
\end{center}
\end{figure}

The above argument implies that it is not necessary to integrate the rate equation from $x = 0$, but only from the freeze-out time $x_f$ to some large time, which we take to be $x = 1000$.
Since we do not know {\it a priori} at what temperature the DM candidate will freeze out we take the following approach that both finds the freeze-out temperature and gives the correct solution to the density evolution equation today.  We start the integration at a temperature $x_i$ that is generally far below the freeze-out temperature at (i.e. $x=30$) and calculate $Y_{\infty}$.  We then step back the starting temperature and again calculate $Y_{\infty}$.  This process is repeated until there are two consecutive values for $Y_{\infty}$ which are within a certain precision from each other defined by the parameter: 
\begin{equation}
\frac{Y_{\infty}(i)-Y_{\infty}(i-1)}{Y_{\infty}(i)}\le \verb|eps_ode|, \nn
\end{equation}
where $i$ marks a particular starting temperature $x_i$. 
Fig.~\ref{fig:ode_int} shows an example calculation of the freeze-out temperature.  Starting with a large $x_i$ the code continues to step back until two consecutive values of $Y_{\infty}$ are within \verb|eps_ode| from each other.  This method of calculating relic density is numerically more stable and faster than starting at $x =1,$ since precision issues of small $dY/dx$ are avoided. 

\subsection{General density evolution equation}

A generic DM model can in principle contain any number of particles and types of interactions.  To calculate the relic abundance in such a model, one needs to track the densities of all DM candidates as well as all possible co-annihilation partners $(\chi_{1},...,\chi_{n})$~\footnote{We will refer to the particles in this list as a whole as DM particles.}, while the rest of the particles are assumed to be in thermal equilibrium or have already decayed to $\chi_1 ... \chi_n$ \footnote{For most practical examples the ``rest of the particles'' simply refer to to SM particles.}.

In the most general case, one needs to categorize the interactions that will affect the dark matter particle densities, as only the interactions which change the number of dark matter particles are relevant.  We can classify them into three different groups: 

\begin{itemize} 
\item DM pair annihilations into SM particles: $\chi_{i}\chi_{j}\to SM$.
\item DM particle scattering off of a thermal background particle: $\chi_{i}X\to\chi_{j}Y$.
\item Decays of one DM species to another.  
\end{itemize}

Next, with all the interactions categorized we can write down the set of density evolution equations:
\begin{eqnarray} \label{eq:coupled_eqns_gen}
\frac{dn_{\chi_{i}}}{dt} &=& - 3Hn_{\chi_{i}} - \sum_{j} (1+\delta_{ij})\langle\sigma(\chi_{i}\chi_{j}\leftrightarrow SM)|v|\rangle\left[n_{x_{i}}n_{x_{j}}-n_{x_{i}}^{EQ}n_{x_{j}}^{EQ}\right] \nonumber \\
 &-& \sum_{j\ne i}\langle\sigma(\chi_{i}X\leftrightarrow\chi_{j}Y)|v|\rangle \left[n_{\chi_{i}} - \frac{n_{\chi_{i}}^{EQ}}{n_{\chi_{j}}^{EQ}}n_{\chi_{j}}\right]n_{\chi_{i}}^{EQ} \nonumber \\
 &-& \sum_{j,k,l}(1+\delta_{ij}-\delta_{ik}-\delta_{il})\langle\sigma(\chi_{i}\chi_{j}\leftrightarrow\chi_{k}\chi_{l})|v|\rangle \left[n_{\chi_{i}}n_{\chi_{j}} - \frac{n_{\chi_{i}}^{EQ}n_{\chi_{j}}^{EQ}}{n_{\chi_{k}}^{EQ}n_{\chi_{l}}^{EQ}}n_{\chi_{k}}n_{\chi_{l}}\right] \nonumber \\
 &-& \sum_{j\ne i}\langle\Gamma(\chi_{i}\to\chi_{j}X)\rangle n_{\chi_{i}} + \langle\sigma(\chi_{j}X\to\chi_{i})|v|\rangle n_{\chi{j}}n_{\chi_{j}}^{EQ} \nonumber \\
 &+& \sum_{j\ne i}\langle\Gamma(\chi_{j}\to\chi_{i}X)\rangle n_{\chi_{j}} - \langle\sigma(\chi_{i}X\to\chi_{j})|v|\rangle n_{\chi_{i}}n_{\chi_{i}}^{EQ}, \label{eq:density_evolution}
\end{eqnarray}

\noindent where $n_{\chi_{i}}$ is the number density of the $i^{th}$ dark matter species, $n_{\chi_{i}}^{EQ}$ is the number density of the $i^{th}$ dark matter species in thermal equilibrium defined in Eq.~\ref{eq:Neq}.  The indices $j$, $k$, and $l$ in Eq. \ref{eq:coupled_eqns_gen} run over all of the DM species.  The coefficient for each term in Eq. \ref{eq:density_evolution} is either a thermally averaged cross section or thermally averaged decay rate.  Both terms are similarly constructed, with thermally averaged annihilation section defined in Eq. \ref{eq:sigv} and decay rate being \begin{eqnarray}
\langle\Gamma(\chi_{i}\to SM)\rangle&=& \frac{1}{n_{\chi_{i}}^{EQ}}\int\prod_{k}\frac{d^{3}p_{k}}{(2\pi)^{3}}\frac{1}{2E_{k}}\sum_{\text{all spins}}|\mathcal{M}(\chi_{i}\to\text{SM})|^{2} \nonumber \label{eq:tad} \\
 &\times& \delta^{4}(p_{1}-p_{2}-p_{3})e^{-E_{1}/T},
\end{eqnarray}
where the product is over all the external momenta of the process.

The first term on the right-hand side of Eq.~\ref{eq:density_evolution} corresponds to the DM pair annihilation processes into SM particles.  The second and third lines describe the DM thermal scattering processes.  The final two lines correspond to the decay and creation processes of the unstable DM particles.   The factors containing the Kronecker deltas are designed to properly count the number of interactions \footnote{For self-annihilating particles there would be a factor of $2$ to properly count the interaction rate.  Several papers note that this is counteracted by a factor of $1/2$ to avoid double counting the initial state phase space in the thermally averaged annihilation cross section for identical particles.  Both factors are then canceled and subsequently ignored \citep{Gondolo:2004sc,Griest:1990kh,Srednicki:1988ce}.}. Notice that the term responsible for semi-annihilations is absent from Eq. \ref{eq:density_evolution}, the treatment of which we plan to include in the future versions of our code.  The current version of  \verb|MadDM|  also omits any dark matter decay terms.

\subsection{Multi-Component Dark Matter with no Co-annihilations}
\label{sec:coupled_evolution}

Many scenarios exist in which it is not possible to cast a system of coupled rate equations into a convenient form of Eq. \ref{eq:density_evolution0}. Such scenarios could arise in models with multiple DM particles that do not contain any canonical co-annihilation diagrams (i.e. $\chi_{i}\chi_{j}\leftrightarrow SM$ with $i\ne j$), as well as the thermal background scattering diagrams which are responsible for establishing the thermal equilibrium between the DM species. In this case, the assumptions necessary to cast a system of coupled density evolution equations into Eq. \ref{eq:density_evolution0} are invalid and one is forced to track the density of all the DM species individually.  If we assume that there are no DM decays, we can write down the coupled set of density evolution equations as
\begin{eqnarray} \label{eq:coupled_eqns}
\frac{dn_{\chi_{i}}}{dt} &=& - 3Hn_{\chi_{i}} - \sum_{j} (1+\delta_{ij})\langle\sigma(\chi_{i}\chi_{j}\leftrightarrow SM)|v|\rangle\left[n_{x_{i}}n_{x_{j}}-n_{x_{i}}^{EQ}n_{x_{j}}^{EQ}\right] \\
 &-& \sum_{j,k,l}(1+\delta_{ij}-\delta_{ik}-\delta_{il})\langle\sigma(\chi_{i}\chi_{j}\leftrightarrow\chi_{k}\chi_{l})|v|\rangle \left[n_{\chi_{i}}n_{\chi_{j}} - \frac{n_{\chi_{i}}^{EQ}n_{\chi_{j}}^{EQ}}{n_{\chi_{k}}^{EQ}n_{\chi_{l}}^{EQ}}n_{\chi_{k}}n_{\chi_{l}}\right] \nonumber. 
\end{eqnarray}

Eq. \ref{eq:coupled_eqns} tell us that the conversion rate of dark matter particles can have large effects on thermal evolution of dark matter, if $ \langle\sigma(\chi_{i}\chi_{j}\leftrightarrow\chi_{k}\chi_{l})|v|\rangle  \sim \langle\sigma(\chi_{i}\chi_{j}\leftrightarrow SM)|v|\rangle.$ This condition can be easily arranged in models with multiple particles in the dark sector (see Ref. \citep{Aoki:2012ub} for a recent study of such as scenario). 

For the purpose of illustration, consider the following dark sector:
\begin{eqnarray}
\mathcal{L}_{DM}&\supset& \frac{m_{X_{1}}^{2}}{2}X_{1}^{2}+\frac{m_{X_{2}}^{2}}{2}X_{2}^{2}+\frac{d_{1}}{4}X_{1}^{4} + \frac{d_{2}}{4}X_{1}^{2}X_{2}^{2} + \frac{d_{3}}{4}X_{2}^{4} \nonumber \\
 &+& \frac{\delta_{1}}{2}H^{\dagger}HX_{1}^{2}+\frac{\delta_{2}}{2}H^{\dagger}HX_{2}^{2}. \label{eq:2rxs}
\end{eqnarray}

The two scalars, $X_{1, 2}$, interact with the rest of the SM particles only through the Higgs via the $\delta$ couplings and with each other through the $d$ couplings.  Notice the absence of terms proportional to proportional to $H^{\dagger}HX_1X_2$, which can easily be arranged by introducing discrete symmetries on the dark matter fields. Such terms would induce mixing between the $X_{1}$ and $X_{2}$ states after the Higgs obtains its vacuum expectation value, as well as generate the canonical co-annihilation diagrams between the two DM eigenstates $X_{1}'$ and $X_{2}'$. 
 Both $X_{1,2}$ in the toy model are  stable so there are no decay diagrams to consider.  The $DM\to DM$ scattering diagrams are mediated by the $d_{i}$ couplings where the interesting interaction comes from the term proportional to $d_2$.  
 
The model in Eq.~\ref{eq:2rxs} is an example of a scenario in which the annihilation rate into light states can be of the same order or smaller than the conversion rate of $X_1 \leftrightarrow X_2$. In the parameter space where the conversion is of the same order as the annihilation rate, the relic density calculation can not proceed according to the rule of Eq. \ref{eq:density_evolution0}. Thus, it is necessary to solve to full set of coupled differential equations of Eq. \ref{eq:coupled_eqns}.
 
To show this numerically we choose the following parameter space point.
\begin{equation}
m_{X_{1}}=200~\text{GeV},\quad m_{X_{2}}=220~\text{GeV},\quad \delta_{1}=1,\quad \delta_{2}=0.01
\end{equation}

\begin{figure}[t]
\begin{center}
\includegraphics*[width=4in]{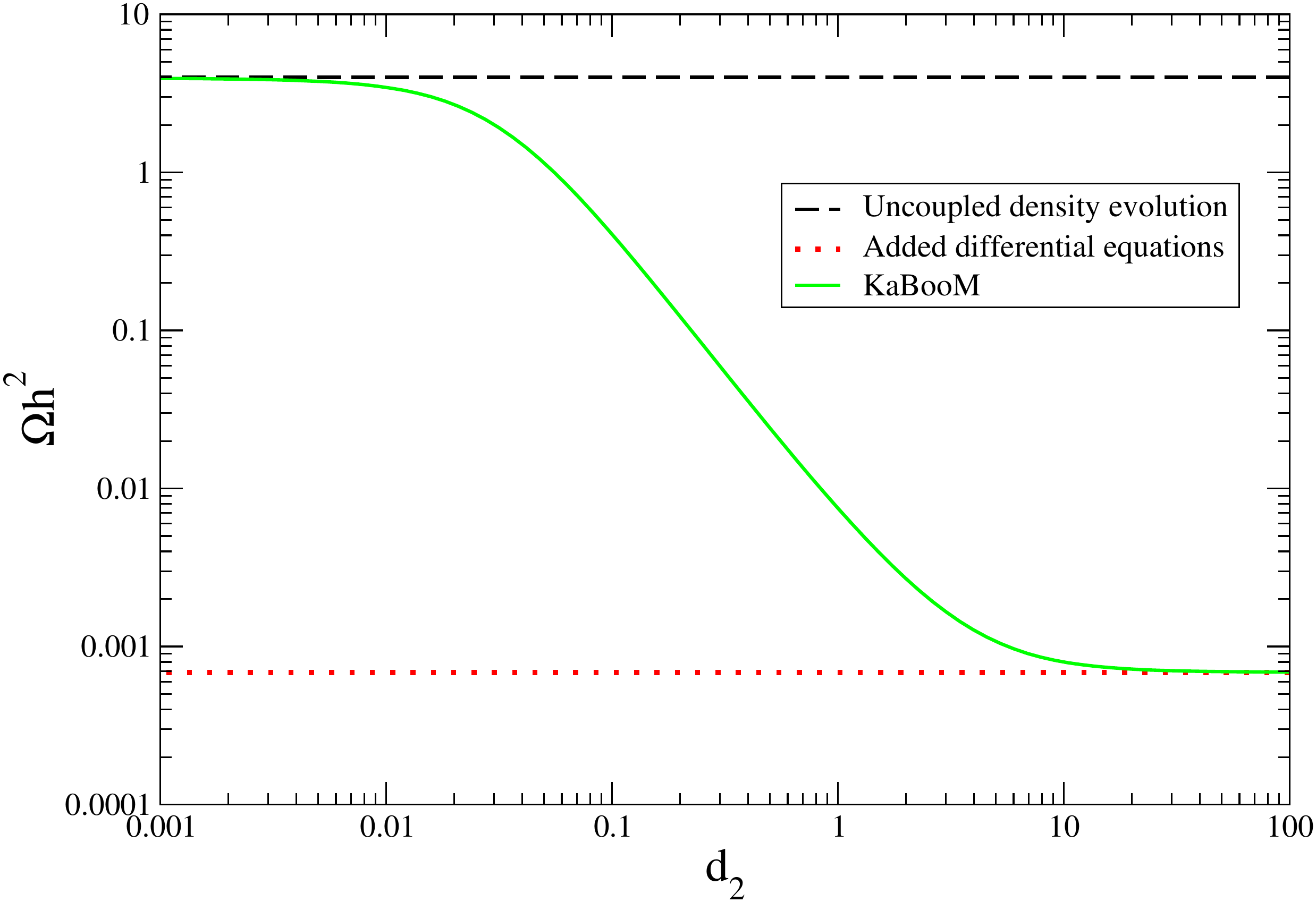}
\caption{Relic density calculation scanned over the strength of the $d_2$ coupling.  With a very small $d_2$ the relic density matches that obtained if the two density evolution equations are solved separately (black line).  Larger $d_2$ couplings give a relic density that is closer to the canonical co-annihilation result that assumes strong thermal scattering interactions (red line). The green line is the result of solving the coupled system of Eq. \ref{eq:coupled_eqns}. For the purpose of the illustration we took $m_{X_1} = 200$ GeV, $m_{X_2} = 220$ GeV, $\delta_{1}=1$, and $\delta_{2}=0.01$.}
\label{fig:d12_plot}
\end{center}
\end{figure}

\noindent Taken by themselves, $X_{1}$ interacts strongly with the Higgs and thus will be underproduced in the early universe while $X_{2}$ is very weakly coupled to the Higgs which would cause over-production. Fig. \ref{fig:d12_plot} shows the calculation of the relic density in this scenario as a function of $d_2$, the coupling which sets the strength of the conversion rate $X_1 \leftrightarrow X_2$. In the limit of $d_2 \rightarrow 0,$ the two dark matter particles are decoupled and the relic density is simply a sum of the relic densities of $X_1$ and $X_2$ which is mainly dominated by the density of $X_{2}$ (solid black line). In the limit of $d_2 \rightarrow \infty$ the high conversion rate keeps $X_{1}$ and $X_2$ in thermal equilibrium and the density evolution is well approximated by Eq. \ref{eq:density_evolution0} (solid red line). However, there is a large region of parameter space in which relic density can take any value in-between the coupled and canonical co-annihilation regime (green line), illustrating that the relic density in this parameter region must be computed using Eq.~\ref{eq:coupled_eqns}.

\section{MadDM Code}

We proceed to discuss the installation of the \verb|MadDM| code as well as the general structure of the code and relevant functions, while we postpone an example \verb|MadDM| script until the Appendix. 

\subsection{Installation}

The current version of \verb|MadDM| is designed to run on most Linux and Mac OS X distributions. Before attempting to run the code, make sure that the following are already installed:
\begin{itemize}
	\item Python 2.6 or 2.7. This is also a requirement for current distributions of \verb|MadGraph 5|. 
	\item \verb|gfortran|.
	\item If the computer is running Mac OS X 10.8, make sure that \verb|make| or \verb|gmake| is installed as they are not included in \verb|Xcode| by default.
	\item \verb|MadGraph 5 v.1.5| or later.
\end{itemize}

Installing \verb|MadDM| is fairly simple and does not require any pre-compilation. 
To install, follow the instructions from 
\begin{verbatim}
       https://launchpad.net/maddm,
\end{verbatim}

\noindent to download the \verb|MadDM| folder inside the main \verb|MadGraph| folder (the path to which we will refer to as \verb|MG| from here on).  We suggest creating a separate \verb|MadGraph| installation dedicated to \verb|MadDM|.

%

\subsection{Running MadDM ``out of the box''}

For the convenience of the user, we have created a user-interface Python script which executes the \verb|MadDM| code. The script can also be used as a skeleton for more elaborate calculations. To run, using the terminal, navigate to the folder in which \verb|MadDM| is installed and run
\begin{verbatim}
       ./maddm.py
\end{verbatim}

We designed the \verb|MadDM| executable script in a way that guides the user through the necessary steps to complete the calculation of dark matter relic density. \verb|MadDM| will prompt the user to select a dark matter model and automatically find a dark matter candidate. The user can then choose whether they would like to include possible co-annihilation channels into the calculation, as well as change the model and \verb|MadDM| parameters.

If the user chooses to change the parameters, \verb|MadDM| will open a MadGraph model parameter card in the \verb|vi| editor. Any modifications made will not affect the parameter card stored in the \verb|MadGraph| models directory. To modify the numerical values of the parameters, you must first press {\bf i} (for ``insert mode''). Do not modify the names of the parameters. To go back to the navigation mode, press {\bf Esc}. To save the changes, first make sure you are in navigation mode. Then press {\bf :} followed by {\bf w}. To exit, press {\bf :} followed by {\bf q}. The two commands can be combined by typing {\bf :wq}. {\bf ctrl + f} and {\bf ctrl + b} can be used navigate the page down and up respectively through the file in the navigation mode. 

After all of the parameters are set, \verb|MadDM| will compile the numerical code and display the resulting relic density.

Alternatively, instead of relying on the ``out of the box'' script, one can write customized Python scripts (see the Appendix for an example) using the \verb|MadDM| classes we discuss in the next Section.

\subsection{Structure of the Code}

The \verb|MadDM| code is divided into two main units: the MadGraph interface (in Python) and the numerical analysis code (in FORTRAN).
The backbone of the Python code is in the \verb|darkmatter.py| and \verb|init.py| files, which are stored in the base directory of \verb|MadDM|. Class definitions, functions which set up the relic density calculation, functions which find a dark matter candidate, generate relevant Feynman diagrams, and handle all the file output are all defined in \verb|darkmatter.py|. These will be discussed in more detail in the following sections. 

The FORTRAN side of the code performs the numerical calculation of the relic density using the FORTRAN matrix elements obtained from MadGraph. These are defined in \verb|relic_coupled.f| (for solving the full set of coupled density evolution equations) and \verb|relic_canon.f| (for the canonical treatment of co-annihilations using the $\langle \sigma_{\rm eff} v \rangle$ proxy).  

The \verb|MadDM| code is structured as follows:
\begin{itemize}
	\item \verb|Projects|: Contains the \verb|MadDM| output for a user specified project. 
	\item \verb|Templates|: Contains skeleton files which are used by the FORTRAN part of the \verb|MadDM| code. The files are divided into several sub-sub-folders:
	\begin{itemize}
		\item \verb|include|: Contains various data used by the FORTRAN numerical module during the calculation of relic density. Files named \verb|boson.txt, fermion.txt, g_star.txt|,  \\ and \verb|g_star_S.txt| contain information about the number of dark matter and light degrees of freedom used in the integration of the density evolution equations. The \verb|MadDM_card.inc| file contains parameters which set the precision of the numerical integrators for the FORTRAN code as well as the flags which turn the status message print-outs on and off. See Section \ref{sec:maddm_inc} for more details.
		\item \verb|matrix_elements|: Contains the templates for the FORTRAN version of the MadGraph matrix element output.
		\item \verb|output|: Empty folder. It is used to store the output of test routines. See Section \ref{sec:test_routines} for more details.
		\item \verb|src|: Contains all the \verb|MadDM| FORTRAN code. Different integration methods, including Romberg, Runge-Kutta and Vegas are defined in \verb|integrate.f|. First and third order interpolation routines as well as the polynomial expansions of the Bessel functions are defined in \verb|interpolate.f|. 
				The density evolution equation is defined and solved by routines in \verb|relic_canon.f| and \verb|relic_coupled.f|. Finally, \verb|maddm.f| is the file containing the main function calls which perform the calculation of relic abundance. It also contains calls (commented out by default) to test routines for de-bugging of matrix-elements, cross sections and thermally averaged cross sections. All of the test routines write their output to the \verb|output| folder.
	\end{itemize}
\end{itemize}

Compiling \verb|maddm.f|, by typing \verb|make| in the base folder of a user's project produces an executable \verb|maddm.x|. This is the file which \verb|MadDM| uses to calculate relic density. Please note that at this stage \verb|maddm.x| is a standalone \verb|MadDM| executable which does not require any additional MadGraph component to run.

\subsection{The darkmatter python class}

We designed the \verb|MadDM| code in a modular way, so that it can be incorporated into any Python script. Here we give a detailed description of the base \verb|darkmatter| class as well as the user interface functions from \verb|init.py| which can be used to construct user-specific Python interfaces for the \verb|MadDM| code. In the following list we label the functions as

\vspace{5 mm}
 \verb| <class name>: <return values>: <function name> ( <arguments> )|.

\vspace{5 mm}
We use the convention of a \verb|'_'| prefix to indicate that a variable name is a \verb|darkmatter| class member.  The \verb|darkmatter| class inherits the native MadGraph classes \verb|base_objects| and \verb|Particle|, allowing for easier implementation of the particle properties.  The relevant member functions are:

\begin{itemize}
	\item \verb|init: [new_proj, model_name, project_name]: | \\
		\verb|  initialize_MadDM_session(print_banner = False)|: This function starts the user interface of the \verb|MadDM| program. It asks the user to input the dark matter model name and checks it against the available dark matter models. It follows with the input of the folder name (subfolder to \verb|Projects|) in which to write all the \verb|MadDM| output. If the project already exists, the function prompts the user to overwrite it. The \verb|print_banner| argument is used to determine whether to print the \verb|MadDM| logo at the start of the program. The default value is False. The return values of the function are a string name of the model and a string name of the project folder with respect to \verb|Projects|, as well as the logical flag which determines whether the project already exists. If the user wishes to hard-code the project name and model, this function can be omitted. 
			
	\item \verb|darkmatter: darkmatter_object: darkmatter()|: Creates an instance of a darkmatter class object. All of the functions below act on an instance of a \verb|darkmatter| class. A \verb|darkmatter| object contains all relevant information about the dark matter candidate, co-annihilation candidates, and the technical details about the \verb|MadDM| project. Here we give an overview of a subset of relevant data members of a \verb|darkmatter| class (all of which are lists):
	\begin{itemize}
      		\item \verb|self._dm_particles|: Contains a list of \verb|Particle| objects of potential dark matter candidates. Information about each particle can be accessed with a simple \verb|print| command (i.e. \verb|print dm._dm_particles[0]|).
      		\item \verb|self._coann_particles|: Contains a list of \verb|Particle| objects of potential dark matter co-annihilation partners.
      		\item \verb|self._dm_names_list|: Contains the list of names of potential dark matter candidates in the \verb|MadGraph 5| user input format. For instance, \verb|n1| is the lightest neutralino of the MSSM etc.
      		\item \verb|self._dm_antinames_list|: Same as \verb|self._dm_names_list| but for anti-particles.
      		\item \verb|self._bsm_particles|: Contains a list of all particles in a model with a Particle Data Group (PDG) code higher than $25$.
      		\item \verb|self._bsm_masses|: Contains a list of numerical values for the masses of all the BSM particles contained in \verb|self._bsm_particles|.
      		\item \verb|self._bsm_final_states|: Contains a list of possible BSM final state particles (in \verb|Particle| object form) that the dark matter candidate can annihilate into. 
      		\item \verb|self._projectname|: Stores the name of the current project.
      		\item \verb|self._projectpath|: Stores the location of the project folder.
		\item \verb|self._paramcard|: Contains the location of the \verb|param_card.dat| file used in the project.
		\item \verb|self._modelname|: Contains the name of the MadGraph 5 model used in the project.
	\end{itemize}

	\item \verb|darkmatter: None: init_from_model | \\
	    \verb|(model_name, project_name, new_proj = True)|: This function initializes the member variables of the \verb|darkmatter| object using the model \verb|model_name|.  The name of the model should correspond to a subfolder in the \verb|models| directory of \verb|MadGraph|. \verb|project_name| is the name of the folder within \verb|/Projects| where the output of \verb|MadDM| will be stored. \verb|init_from_model| will create the \verb|Projects/project_name| folder and copy the contents of the \verb|/Template| folder to it.  If \verb|new_proj = True| the code will overwrite any existing project with the name \verb|project_name|. The function returns no value.

	\item \verb|darkmatter: None: FindDMCandidate(dm_candidate='', prompts = True)|: \\ Based on the criteria that the particle is non-Standard Model and stable,  this function finds the dark matter candidates in a model defined by the \verb|_modelname| member. It returns no value, but it sets the member variables of a \verb|darkmatter| object. Information about the dark matter candidates is stored in the 
	\verb|_dm_particles| member array. It also adds all the BSM particles to the \verb|_bsm_particles| member array. With the default options, the function will prompt the user to change the model parameter card before determining the dark matter candidates. It will also prompt the user to enter the dark matter candidate by hand. If no user input is provided, the function will automatically determine the dark matter candidate from the default parameter card in the \verb|MadGraph| model directory and print out the properties of the resulting dark matter candidate. The prompts can be turned off by setting the \verb|prompts| flag to \verb|False|. The user can also hard-code the dark matter candidate by specifying the \verb|dm_candidate| input parameter. 	
	
	\item \verb|darkmatter: None: Findco-annihilationCandidates| \\ \,\,\,\,\verb| ( prompts = True, coann_eps = 0.1)|: Determines the co-annihilation candidates \\ based on the mass splitting criteria
	\be
		\verb|coann_eps| \equiv \frac{|m_{\chi_0} -m_{ \chi_1}|}{m_{\chi_0}}, \label{eq:eps}
	\ee
	where $\chi_{0}$ is the dark matter candidate with mass $m_{\chi_0}$ and $\chi_1$ is the co-annihilating partner with mass $m_{\chi_1}$. The default is $\verb|coann_eps| = 0.1$, and can be passed to the function as a parameter. If the mass splitting is less than $\verb|coann_eps|$ and a particle is a BSM particle (i.e. PDG $> 25$), it is added to the \verb|_dm_particles| member array. If \verb|prompts = True| the function will prompt the user to enter the co-annihilation candidates by hand. If no input is provided, the co-annihilation candidates are determined automatically, based on the masses in the default parameter card. If the user inputs the co-annihilation partners by hand, the mass splitting criteria is omitted. The default value for \verb|prompts| will also result in a print-out of the co-annihilation candidates. The prompts and messages can be turned off by setting the \verb|prompts| flag to False. 
	
	The manual selection of the co-annihilation partners is particularly useful in a parameter scan where the range of particle masses is large. See Section \ref{sec:paramscan} for more details.

	\item \verb|darkmatter: None: GenerateDiagrams()|: This function generates the $2 \rightarrow 2$ dark matter annihilation diagrams. The initial states are formed from all valid combinations of particles stored in the \verb|_dm_particles| member array, whereas the final states are all Standard Model particles in addition to particles stored in the \verb|_bsm_final_states| member array.

	\item \verb|darkmatter: None: CreateNumericalSession(prompts = True)|: Outputs the FORTRAN files containing the matrix elements and necessary numerical code to perform the relic density calculation. It also compiles the code and creates the executable \verb|maddm.x| within the \verb|Projects/<proj_name>| folder. If \verb|prompts| is set to True, it will print out status messages and a list of annihilation processes included in the calculation. The function will also prompt the user to edit the \verb|maddm_card.inc| file (see Section \ref{sec:maddm_inc}). If \verb|prompts = False| the default \verb|maddm_card.inc| from the \verb|Templates/include| directory will be used. 
	
	If the user wishes to turn off the prompts, changes to the \verb|maddm_card.inc| can still be made in the \verb|include| subfolder of the \verb|Projects/<proj_name>| directory. Note however that the numerical code must be recompiled for changes to take effect.

	\item \verb|darkmatter: double: CalculateRelicAbundance() |: \\This function executes \verb|maddm.x| in the project folder \verb|Projects/<proj_name>|. It returns a numerical value for $\Omega h^2$.  This function can not be executed before calling \\ \verb|GenerateDiagrams| and \verb|CreateNumericalSession|. 
	
	\item \verb|darkmatter: double: GetMass(pdg_id)|:  Returns the numerical value for the mass of a particle with a PDG code \verb|pdg_id|.
	
	\item \verb|darkmatter: double: GetWidth(pdg_id)|: Returns the numerical value for the width of a particle with a PDG code \verb|pdg_id|.
	
	\item \verb|darkmatter: None: ChangeParameter(parameter_name, parameter_value)|: \\ This function changes the numerical value of an entry in the \verb|param_card.dat| file (specified by the \verb|self._param_card| member variable). \verb|parameter_name| is a string and must correspond to the parameter label in \verb|param_card.dat|. For instance, \\ \verb|ChangeParameter('Mt', 175.0)| will change the numerical value of the top mass in the parameter card to $175.0\, { \rm GeV}$. 
	
	Note that default settings in the \verb|include/maddm_card.inc| can cause issues with output formatting in a parameter scan.  if you wish to eliminate the status messages from the FORTRAN side of the code during the parameter scan, you must set the \verb|print-out| flag in the \verb|include/maddm_card.inc| file to \verb|.false.|, and then recompile \verb|maddm.x| by typing \verb|make| in the base project directory.
	
	\item \verb|darkmatter: string : GetProjectName()|: Returns the name of the folder in the \verb|Projects| directory where the \verb|MadDM| output is stored and checks whether or not the project folder already exists.
	
	\item \verb|darkmatter: string: Convertname(name)|: Converts particle names (e.g. e+ $\rightarrow$ ep) so that they can be used as names for files and folders.
	
	\item \verb|darkmatter: None: init_param_scan(name_of_script)|: Creates the Python script which executes a parameter scan. The function takes in one parameter, \verb|name_of_script| which is the name of the file the parameter scan script should be saved to in the \verb|Projects/<proj_name>| folder. The function will edit a default template script in the \verb|vi| editor.  Make sure to save the script upon editing it by typing \verb|:wq| in the navigation mode of \verb|vi|. The script will be automatically executed upon exiting \verb|vi|. 
	
\end{itemize}

\subsection{A Remark about Parameter Scans} \label{sec:paramscan}

 When performing parameter scans with \verb|MadDM|, it is important to be aware of the algorithm the code uses to calculate relic density, so as to avoid possible bugs in the computation. 

One issue that might arise is in models with co-annihilations. If the initial mass splitting between the dark matter and co-annihilating partners is too large when the diagram generation step takes place, the co-annihilation diagrams will not be included in the calculation. The parameter scan takes place only upon the compilation of the numerical code, and it is not possible for \verb|MadDM| to add diagrams midway through the scan without recompiling the entire numerical module. In order to bypass this issue select co-annihilation candidates by manually selecting which particles you would like to consider at the diagram generation step. Alternatively, one can also initialize the masses in the \verb|param_card.dat| to values which are within the $\verb|coann_eps|$ of Eq. \ref{eq:eps} and let \verb|MadDM| automatically find the co-annihilation partners.

\subsection{The MadDM include file} \label{sec:maddm_inc}

The \verb|MadDM_card.inc| file contains relevant parameters for the numerical calculation of relic density. These parameters should be set in the pre-compilation step of the calculation and once the numerical code is compiled, they can be modified by editing the \verb|MadDM_card.inc| and recompiling the code. The relevant variables in the file are
\begin{itemize}

\item \verb|print_out|: A logical flag which determines whether the numerical code should print out status messages. It is set to \verb|.false.| by default.

\item \verb|relic_canonical|: A logical flag which switches between the canonical treatment of relic density in Eq. \ref{eq:density_evolution0} and the full, coupled equation treatment of Eq. \ref{eq:coupled_eqns}. It is set to \verb|.true.| (canonical treatment) by default.

\item \verb|calc_taacs_ann_array|: A logical flag which when set to \verb|.true.| will calculate the thermally averaged annihilation cross sections and store them in an array for interpolation during the integration of the density evolution equation.

\item \verb|calc_taacs_dm2dm_array|: In the case of coupled density evolution equations (see Section \ref{sec:coupled_evolution}) this is a logical flag that calculates and stores the thermally averaged annihilation cross section for all of the dark matter conversion processes.

\item \verb|calc_taacs_scattering_array|: Similar to \verb|calc_taacs_dm2dm_array| but the thermally averaged annihilation cross sections are for DM/SM elastic scattering processes This function is useful only for de-bugging purposes. In principle, if SM elastic scattering processes are present then one should use the canonical co-annihilation formalism as opposed to the coupled density evolution equations. 

\item \verb|eps_ode|: A parameter which sets the precision required from the ODE integrating routines in determination of the freeze-out temperature. The integration of the density evolution equations stops stepping back in $x$ (see Section \ref{sec:boltzprac}) when
\begin{equation}
\frac{Y_{\infty}(i)-Y_{\infty}(i-1)}{Y_{\infty}(i)}\le \verb|eps_ode|, \nn
\end{equation}
where $Y \equiv n/s $ is the dark matter number density per unit entropy.
\item \verb|eps_wij, eps_taacs|: Parameters which set the precision requirements of the Romberg method integrator for the $W_{ij}$ values (see Eq. \ref{eq:wij}), and the thermally averaged cross section. The integration stops when
\be
	\frac{I(i) - I(i-1)}{I(i)}  \le \verb|eps| \nn,
\ee
where $I$ is the value of the integral in question, $i$ is the iteration of the Romberg algorithm and \verb|eps| is the desired precision (i.e. \verb|eps_taacs| or \verb|eps_wij|).

\item \verb|iter_wij, iter_taacs|: The minimum number of Romberg iterations, independent of the required precision. These parameters are particularly useful when considering narrow $s$-channel resonances.  The default values are set to $10$ iterations. 

\item\verb|beta_step_min, beta_step_max|: The minimum and maximum step size for the calculation of $W_{ij}$ values (see Eq. \ref{eq:wij}). The routines that calculate the $W_{ij}$ arrays with respect to beta will automatically adapt the step sizes between the minimum and maximum values specified here. 

\end{itemize}

\subsection{Using the test routines} \label{sec:test_routines}

For debugging purposes, we have added a number of test routines to the \verb|MadDM| code: 

\begin{itemize}
	\item \verb|bessel_check (min_x, max_x, nsteps, logscale)|: prints out the numerical values of the Bessel function approximations used in the relic density evolution equation. \verb|x_min| and \verb|x_max| define the range of $x \equiv m/T$ values, while \verb|nsteps| defines the number of output points. \verb|logscale| is a flag which can take values of 0 (linear scale output) and 1 (log scale output). 
	\item \verb|dof_check(min_temp, max_temp, nsteps, logscale)|: same as \verb|bessel_checl| except for the interpolation of the functions $g_*(T)$, the number of relativistic degrees of freedom, and $g_S(T)$, the number of degrees of freedom which contributes to the entropy density.
	\item \verb|Wij_check()|:prints out the values of $W_{ij}$ (see Eq. \ref{eq:wij}) as a function of the     center of mass energy into a text file. 
	\item \verb|taacas_check(min_x, max_x, nsteps, logscale)|: prints out the values of the velocity averaged cross section $\langle \sigma v \rangle$ as a function of $x \equiv m / T$ into a text file. \verb|min_x| and \verb|max_x| define the range of $x$ values and \verb|nsteps| defines the number of data points to output. The flag \verb|logscale| can be used to output the  $\langle \sigma v \rangle$ values on the log scale ($1$ for log scale, $0$ for linear). 
	\item \verb|odeint_check(new_or_old)|: prints out the values of $Y \equiv n/s$  as a function of $x$. The flag \verb|new_or_old| defines whether the calculation should be performed in the canonical formulation (see Eq. \ref{eq:density_evolution0}) (0) or whether the full set of coupled equations (see Eq. \ref{eq:coupled_eqns}) should be considered (1). 
	      \item  \verb|cross_check_scan_all(xinit_min, xinit_max,| \\ 
		\verb|nsteps, p_or_E, logscale)|: prints out the cross section $\sigma$ summed over all the relevant dark matter annihilation processes as a function of initial state energy or momentum in the center of mass frame. \verb|nsteps| determines how many data points to output. \verb|p_or_E| is a flag which determines whether the cross section should be output as a function of initial state momentum (0) or energy (1). As before, \verb|logscale| determines whether the output should be log scale (1) or linear (0). The parameters \verb|xinit_min| and \verb|xinit_max| define the range of momentum/energy (depending on the value of the \verb|p_or_E| flag) for the cross section calculation. 
	\item \verb|cross_check_scan(dm_i, dm_j, process_k, xinit_min,| \\
		\verb| xinit_max, nsteps, p_or_E, logscale) |:  prints out the cross section $\sigma$ for an individual process as a function of initial state energy or momentum. The parameters \verb|dm_i| and \verb|dm_j| select which particles to consider in the initial state. For instance, \verb|dm_i = 1| and \verb|dm_j = 2| will select the processes in which dark matter co-annihilates with the next lightest BSM particle. \verb|process_k| selects a particular annihilation channel to be calculated. \verb|xinit_min|, \verb|xinit_max| define the range of values of momentum or energy (depending on the \verb|p_or_E| flag for the cross section calculation). 
	\item \verb|cross_check_process(dm_i, dm_j, process_k, x1init, x2init, p_or_E)|:\\ same as the previous function, but for a single momentum/energy defined by \verb|x1init| and \verb|x2init|.
	\item \verb|matrix_element_check_all(x1init, x2init, p_or_E)|: same as the \\ \verb|cross_check_scan_all|  function except the numerical value of the matrix element is printed out instead of the cross section.
	\item \verb|matrix_element_check_process(dm_i, dm_j, process_k,  x1init, x2init,| \\ \verb| p_or_E)|: same as the  \verb|cross_check_process| function except the numerical value of the matrix element is printed out.

\end{itemize}

The test routines are automatically added to every project in the \verb|src\tests.f| file. In addition, calls to the test routines are included as commented-out parts of the \verb|maddm.f| file. To use the test routines, simply navigate to the \verb|src| folder within your project directory and edit the \verb|maddm.f| file. Uncomment the test routines you wish to use and save the changes. Navigate back to the project folder and recompile the \verb|MadDM| code by executing \verb|make|. 

The output of the test routines is stored in the text files in the \verb|output| folder within your project directory. 

\subsection{Interfacing MadDM with Mass Spectrum Generators}

Many models of dark matter involve more than one particles in the BSM sector and complex mass spectra.  Dedicated pieces of software are often available to translate the model parameters into the masses of physical states and decay rates to light particles. The most notable examples come from supersymmetry where a myriad of tools such as SUSPECT~\citep{Djouadi:2002ze},  SUSYHIT~\citep{Djouadi:2006bz}, etc.  
have been developed. 

\verb|MadDM| is easily interfaced with any model specific spectrum or decay width code which can produce a Les Houches formatted \verb|MadGraph| parameter card. After producing a \verb|maddm.x| executable, simply copy the parameter file from the spectrum generator as the \verb|param_card.dat| in the \verb|/Cards| subfolder of your project.  Upon re-running \verb|maddm.x|,  the code will read the new parameters and output the corresponding value for relic density.

\section{Validation}

We performed detailed validation of the \verb|MadDM| code in a wide range of dark matter models against results obtained with \verb|micrOMEGAs| 2.4.5 \citep{Belanger:2010gh}. We chose the models specifically to test the non-standard scenarios such as resonant annihilations and co-annihilations. For models which combine these features we compare the \verb|MadDM| calculation of relic density in the simplest forms of the extra dimensional and supersymmetric theories to knows results. 

\label{sec:val}

\subsection{Real Scalar Extension of the Standard Model (RXSM)}

The first and simplest model we consider is the Real Scalar Extension of the Standard Model (RXSM) \citep{Barger:2007im}. The model consists of an additional scalar singlet field, which couples to the Standard Model only through the Higgs:

\begin{eqnarray}
V &=& {m^2\over 2} H^\dagger H+{\lambda\over 4}(H^\dagger H)^2+{\delta_1\over 2} H^\dagger H S+{\delta_2\over 2} H^\dagger H S^2\nn\\
&+&\left({\delta_1 m^2\over 2 \lambda}\right) S + {\kappa_2 \over 2} S^2+{\kappa_3\over 3} S^3+{\kappa_4 \over 4} S^4,
\label{eqn:hpot}
\end{eqnarray}
where $S$ is the scalar-singlet field. A discrete $Z_2$ symmetry stabilizes $S,$ making it a dark matter candidate. For the purpose of code validation we omit the details of spontaneous symmetry breaking and leave the mass of $S$ a free parameter. Thus, the only relevant parameters for the relic density calculation are the mass $m_S$ and the coupling of $S$ to $H$ which we denote as $\delta$. The $\kappa_n$ couplings only contribute to the masses and are omitted from the analysis.

\begin{figure}[tbp]
\begin{center}
\includegraphics[angle=0,width=0.16\textwidth]{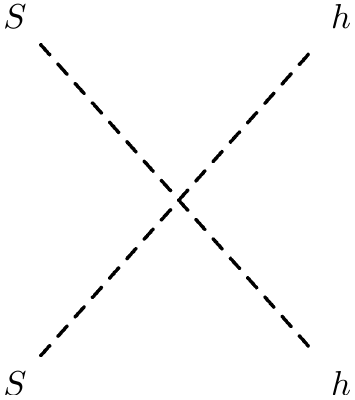}~~~
\includegraphics[angle=0,width=0.185\textwidth]{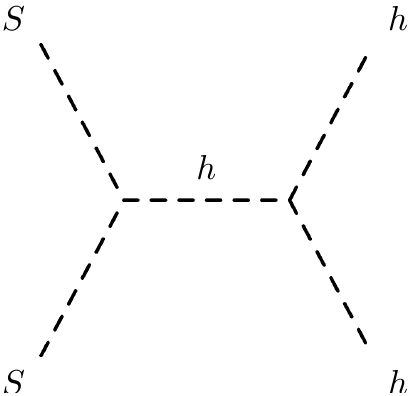}~~~
\includegraphics[angle=0,width=0.185\textwidth]{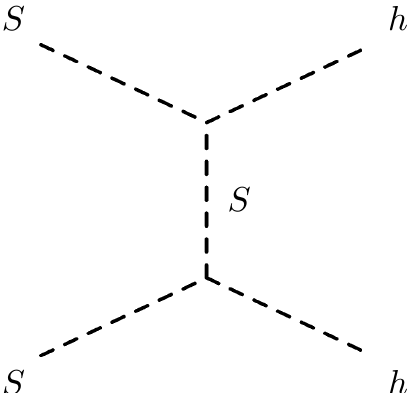}~~~
\includegraphics[angle=0,width=0.185\textwidth]{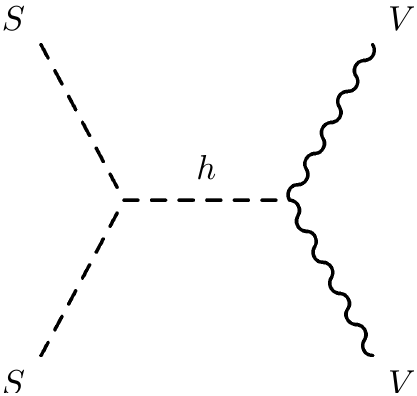}~~~
\includegraphics[angle=0,width=0.185\textwidth]{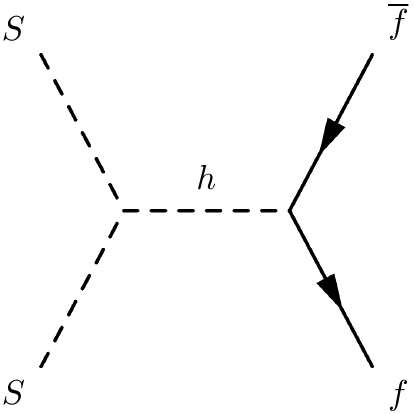}
\caption{Annihilation processes that contribute to the thermally averaged cross section.  Taken from Ref. \citep{Barger:2007im}.}
\label{fig:annfd}
\end{center}
\end{figure}

\begin{figure}[tbp]
\begin{center}
\includegraphics[scale = 0.65]{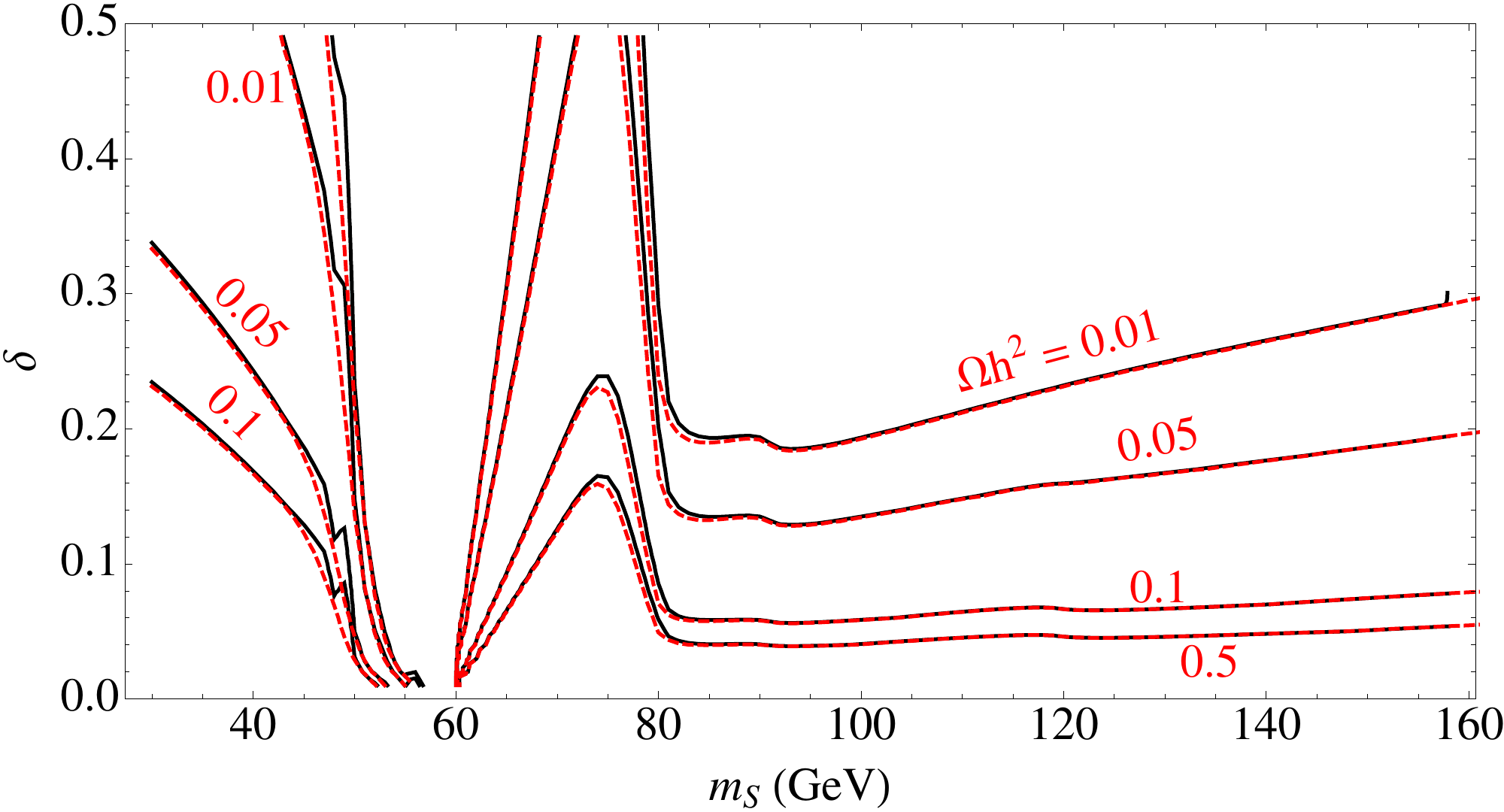}
\caption{Comparison between the relic density results obtained with MadDM and micrOMEGAs. The black, solid contours are the results obtained from MadDM. The red, dashed contours are obtained with micrOMEGAs. The agreement is within $5 \%$ for most of the parameter space, except very close to the resonance where precision issues are expected to arise.}
\label{fig:comp_rxsm}
\end{center}
\end{figure}

The RXSM model is useful for \verb|MadDM| validation as it contains a resonant $SS \rightarrow h \rightarrow SM$ channel in the list of annihilation diagrams shown in Fig.~\ref{fig:annfd}.  Fig.~\ref{fig:comp_rxsm} shows the result  for the relic density calculation with an excellent agreement between our code and \verb|micrOMEGAs| \footnote{The relic density calculation is very sensitive to the running $b$ mass and the running $b$ yukawa coupling in the resonant region. For the purpose of the comparison, we use a fixed $m_b = 4.7 \GeV$ and a vacuum expectation value (vev) of $v = 246 \GeV$ in both codes.}, except in the region where the mass of dark matter $m_h / 2 - 5\, {\rm GeV} < m_S  < m_h / 2$, where we find a significant discrepancy of $O(1)$. 

The large discrepancy is a result of known precision issues when dealing with ultra-narrow resonances close to threshold. The calculation of thermally averaged cross section in this region involves a convolution of a narrow Breit-Wigner function with a velocity distribution which can be approximated as $\beta^2 e^{-\beta^2/ 2\beta_0^2}$. If the resonance is close to and above threshold, the integral involves a numerically complex saddle point as the slopes of both the Breit-Wigner and the velocity distribution can be extremely steep. 
Note that the discrepancy does not occur for $m_S > m_h / 2$ even close to the resonance. Since the resonance is below threshold in this case, the dark matter distribution convolves only the smooth tail of the Breit-Wigner function and no saddle point numerical issues arise. 

In order to improve the numerical accuracy when calculating relic density in models with very narrow resonances close to threshold ($i.e.$ Higgs), we recommend to increase the value of \verb|iter_wij| and \verb|iter_taacs| in the \verb|maddm_card.inc| file. Note that for any change of this file to take effect you must recompile the Fortran module of the \verb|MadDM| code. 

\subsection{Complex Scalar Extension of the Standard Model (CXSM) } \label{sec:cxsm}

Models with additional scalars offer more complex BSM sectors. In this section we consider a Complex Scalar Extension of the Standard Model (CXSM) \citep{Barger:2008jx,Barger:2010yn}  as a benchmark for comparisons between \verb|micrOMEGAs|  and \verb|MadDM|. 

We begin by considering a general renormalizable scalar potential obtained by the addition of a complex scalar singlet to the SM Higgs sector:
\bea
V(H,\cs) &=& \frac{m^2}{2}H^\dagger H+\frac{\lambda}{4}(H^\dagger H)^2+\left(\frac{|\delta_{1}|e^{i \phi_{\delta_1}}}{4} H^\dagger H \cs + c.c.\right)+\frac{\delta_2}{2} H^\dagger H |\cs|^2\nn\\
&+&\left(\frac{|\delta_{3}|e^{i \phi_{\delta_3}}}{4} H^\dagger H \cs^2+h.c\right)
+\left(|a_{1}|e^{i \phi_{a_{1}}}\cs+c.c.\right)+\left(\frac{|b_{1}|e^{i \phi_{b_1}}}{4}\cs^2+c.c.\right)\nn\\
&+&\frac{b_{2}}{2}|\cs|^2+\left(\frac{|c_1| e^{i \phi_{c_1}}}{6}\cs^3+c.c.\right)
+\left(\frac{|c_2| e^{i \phi_{c_2}}}{6}\cs|\cs|^2+c.c.\right)\nn\\
&+&\left(\frac{|d_1| e^{i \phi_{d_1}}}{8}\cs^4+c.c.\right)+\left(\frac{|d_3| e^{i \phi_{d_3}}}{8} \cs^{2}|\cs|^{2}+c.c.\right)+\frac{d_2}{4}|\cs|^4,
\label{eq:pot}
\eea

where $H$ is the $SU(2)$ doublet field that acquires a vev:

\be
\langle H\rangle = \left(
\begin{array}{c}
0 \\ v/\sqrt{2}
\end{array}\right) .
\ee

Here, $m^{2}$ and $\lambda$ are the usual parameters of the SM Higgs potential, while we again adopt the SM vacuum expectation value of $v=246$ GeV.

A discrete $Z_2$ symmetry on $\cs$ serves to eliminate all the terms containing odd powers of the field, thus preventing preventing the decay of $\cs$. The remaining scalar potential takes the form:
\bea
V_\mathrm{CXSM}&=&\frac{m^{2}}{2}H^{\dagger}H+\frac{\lambda}{4}(H^{\dagger}H)^{2}+\frac{\delta_{2}}{2}H^{\dagger}H|\cs|^{2}+\frac{b_{2}}{2}|\cs|^{2}+\frac{d_{2}}{4}|\cs|^{4}
\label{eq:u1pot} \\
\nonumber
&+&\left(\frac{|b_{1}|}{4}e^{i\phi_{b_1}}  \cs^{2}+|a_1|\, e^{i\phi_{a_1}} \cs  + c.c.\right) .
\eea
The result is a BSM sector which contains a stable dark matter candidate and a heavier state which dark matter can co-annihilate with, making CXSM an excellent framework for testing the ability of \verb|MadDM| to perform relic density calculations in models with co-annihilations. Relevant annihilation diagrams of CXSM are similar to the ones in Fig. \ref{fig:annfd}, replacing $S$ with $A$. $H_1$ and $H_2$ are also possible in the place of $h$ in Fig.\ref{fig:annfd}.

\begin{figure}[t]
\begin{center}
\includegraphics[scale = 0.65]{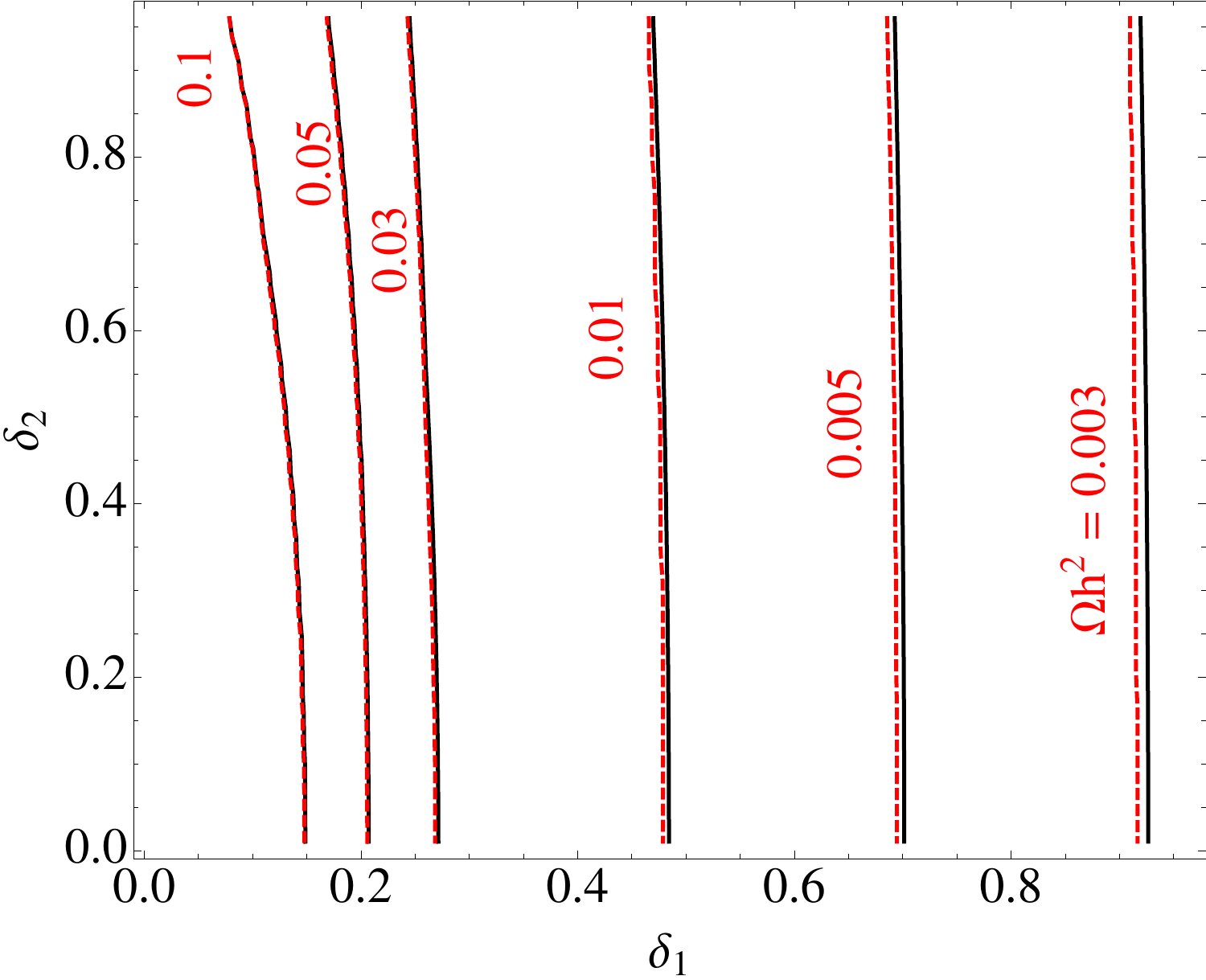}
\caption{Comparison between the relic density results obtained with MadDM and micrOMEGAs in the CXSM model. 
The black, solid curves are contours of constant  $\Omega h^2$, which are obtained from MadDM. 
The red, dashed contours are obtained with micrOMEGAs. 
The agreement is within $5 \%$ for the entire region of the parameter space. 
For the purpose of the comparison, we used $m_{\chi_1} = 400$ GeV, $m_{\chi_2} = 420$ GeV and $\delta_{12} = 0.05$. }
\label{fig:comp_cxsm}
\end{center}
\end{figure}

Fig. \ref{fig:comp_cxsm} shows a numerical comparison between \verb|MadDM| and \verb|micrOMEGAs| in the context of the CXSM model. We find that the results agree within $5 \%$ over a wide range of parameter space. The agreement with \verb|micrOMEGAs| can be further improved by decreasing the \verb|eps_wij, eps_taacs| and \verb|eps_ode| precision parameters of the \verb|MadDM| code. In cases where the dominant annihilation channel is through a narrow $s$-channel resonance, additional handle on precision is the minimum step in dark matter velocity \verb|beta_step_min| (used in the integration of $\langle \sigma v \rangle$ of Eq. \ref{eq:sigv}). Finally, the minimum number of iterations \verb|iter_wij| and \verb|iter_taacs| can be increased to ensure that the possible strongly varying features of cross sections are accurately captured by the Romberg integration routines. 

\subsection{MSSM} \label{sec:mssm}

\scriptsize
\begin{table}
\begin{eqnarray}
\tiny
{\rm \mathbf{SPS 1a}}:& \nonumber \\ \nonumber
{\rm Point:}&  m_0 = 100 \GeV, \,\, m_{1/2} = 250 \GeV, \,\, A_0 = -100 \GeV, \\ &  {\rm tan} \beta = 10, \,\, \mu>0.  \nonumber\\
{\rm Slope}:&m_0 = -A_0 = 0.4 m_{1/2}, \,\, m_{1/2} \,\,  {\rm varies.} \nonumber\\ \nonumber\\
{\rm \mathbf{SPS 1b}}:& \nonumber \\\nonumber
{\rm Point:}&  m_0 = 200 \GeV, \,\, m_{1/2} = 450 \GeV, \,\, A_0 = 0, \\ &  {\rm tan} \beta = 30, \,\, \mu>0.  \nonumber\\\nonumber
{\rm Slope}:&m_0 = -A_0 = 0.4 m_{1/2}, \,\, m_{1/2} \,\,  {\rm varies.} \nonumber\\ \nonumber\\\nonumber
{\rm \mathbf{SPS 2\, }}:& \nonumber \\\nonumber
{\rm Point:}&  m_0 = 1450 \GeV, \,\, m_{1/2} = 300 \GeV, \,\, A_0 = 0 \GeV, \\ &  {\rm tan} \beta = 10, \,\, \mu>0.  \nonumber\\\nonumber
{\rm Slope}:&m_0 = 2 m_{1/2} +850  \GeV, \,\, m_{1/2} \,\,  {\rm varies.} \nonumber\\ \nonumber\\\nonumber
{\rm \mathbf{SPS 3\, }}:& \nonumber \\\nonumber
{\rm Point:}&  m_0 = 90 \GeV, \,\, m_{1/2} = 400 \GeV, \,\, A_0 = 0 \GeV, \\ &  {\rm tan} \beta = 10, \,\, \mu>0.  \nonumber\\ \nonumber\\\nonumber
{\rm \mathbf{SPS 4\,}}:& \nonumber \\ \nonumber
{\rm Point:}&  m_0 = 400 \GeV, \,\, m_{1/2} = 300 \GeV, \,\, A_0 = 0 \GeV, \\ &  {\rm tan} \beta = 50, \,\, \mu>0.  \nonumber\\ \nonumber\\\nonumber
{\rm \mathbf{SPS 5\, }}:& \nonumber \\\nonumber
{\rm Point:}&  m_0 = 150 \GeV, \,\, m_{1/2} = 300 \GeV, \,\, A_0 = 0 \GeV, \\ &  {\rm tan} \beta = 10, \,\, \mu>0.  \nonumber \\\nonumber
& {\rm at \, GUT\, scale: } M_1= 480 \GeV, \,\, M_2 = M_3 = 300, \nonumber\\ \nonumber\\ \nonumber
{\rm\mathbf{SPS 9\, }}:& \nonumber \\
{\rm Point:}&  m_0 = 450 \GeV, \,\, m_{{\rm aux}} = 60 \TeV, \nonumber \\  &{\rm tan} \beta = 10, \,\, \mu>0.  \nonumber\\ \nonumber
{\rm Slope}:&m_0 = 0.0075 m_{{\rm aux}}, \,\, m_{{\rm aux}} \,\, {\rm varies.} \nonumber
\end{eqnarray}
\caption{SPS points}
 \label{table:sps}
\end{table}

\large

One of the more complicated models which offers a dark matter candidate is the MSSM. The parameter space of the MSSM is vast enough to require a dedicated study, and is beyond the scope of this paper. We instead choose to test \verb|MadDM| only against several Snowmass Points and Slopes (SPS) parameter choices \citep{Allanach:2002nj}. The benchmark points are characterized in Table \ref{table:sps}.

For the purpose of comparison we used the SUSYHIT \citep{Djouadi:2006bz} package to calculate the MSSM mass spectra and decay rates in \verb|MadDM|. Table \ref{table:mssmcomp} shows the results against \verb|micrOMEGAs| v.3.6.7. Note that in \verb|MadDM| we used the MSSM implementation from the FeynRules \citep{Degrande:2011ua}  website, rather than the default MSSM implementation provided in \verb|MadGraph|. We obtain good agreement for all tested SPS points, except in the case of SPS 4 where we find a discrepancy of  $25 \%$. The SPS4 point is in the, so called, ``resonance region'' of the MSSM parameter space, where the resonant annihilation through the Higgs dominates the velocity averaged cross section. We find that in this region, the calculation of relic density is highly sensitive to the details of running $b$ masses, the problem which is beyond the scope of this paper. 
\\

\begin{table}[htb]
\begin{center}
\begin{tabular}{|c|c|c|c|c|c|c|}
\hline
SPS Pt. & \verb|MadDM| & \verb|micrOMEGAs| & Difference & Description \\ \hline 
SPS1a & 0.197 & 0.195 & $1\%$  &neutralino DM \\
SPS1b & 0.390 & 0.374 & $ 5\%$& coan. with stau \\
SPS2 & 7.914 & 7.860 & $ 1 \%$ & focus pt, higgsino DM\\
SPS3 & 0.118 & 0.116 & $ 2 \% $ &conn. with sleptons \\
SPS4 & 0.0596 & 0.0474 & $ 22 \%$&resonance region \\
SPS5 &  0.332 & 0.338 & $-2\%$&neutralino DM \\
SPS9 & 0.00111 & 0.00117 & $-4 \%$&AMSB, coan. w/ chargino \\
\hline

\end{tabular}
\caption {Values for $\Omega h^2$ for different $SPS$ points of MSSM between micrOMEGAs and MadDM. }
\label{table:mssmcomp}
\end{center}
\end{table}

\subsection{Minimal Universal Extra Dimensions (MUED)}

\begin{figure}[tbp]
\begin{center}
\includegraphics[scale = 0.7]{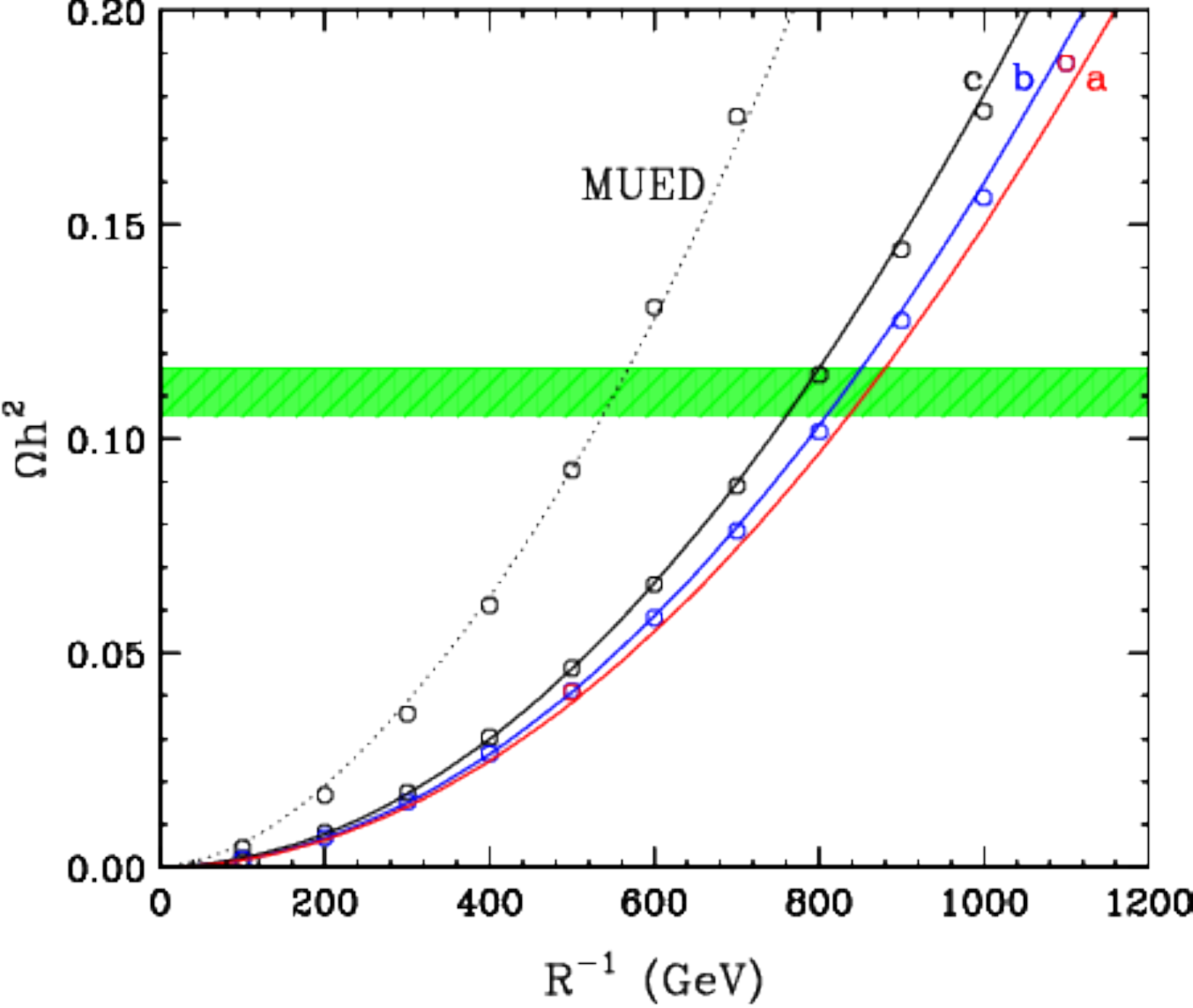}
\caption{Comparison between the relic density results in a MUED model between MadDM and the existing literature. The plot shows the relic density $\Omega h^2$ as a function of the extra dimension radius $R$, the only parameter in the model. }
\label{fig:comp_mued}
\end{center}
\end{figure}

Models of extra dimensions offer a rich particle phenomenology which often provides a viable dark matter candidate (given a symmetry which stabilizes the lightest Kaluza-Klein mode). Minimal Universal Extra Dimensions (MUED) is the simplest of such models, 
where the lightest Kaluza-Klein particle (LKP) is the Kaluza-Klein (KK) photon $\gamma_1$. 
Dark matter in the context of MUED has been extensively studied in Refs. \citep{Servant:2002hb, Arrenberg:2008wy,Kong:2005hn,Arrenberg:2013paa}. 
For the purpose of \verb|MadDM| validation, we compared the results from our code to the results obtained in Refs. \citep{Servant:2002hb,Kong:2005hn}. 
Fig. \ref{fig:comp_mued} shows comparison of the relic density of the LKP as a function of $R^{-1}$, the inverse of the size of the extra dimension. 
Four curves (both solid and dotted) are reproduced following Ref. \citep{Kong:2005hn}, 
while circles on each curve (with 100 GeV interval on $R^{-1}$) are obtained with our code, following the same assumptions for each case.
The (red) line marked ``a" is the result from considering $\gamma_1 \gamma_1$ annihilation only and 
the (blue) line marked ``b" repeats the same analysis, but uses a temperature-dependent 
$g_\ast$ and includes the relativistic correction to the b-term \citep{Kong:2005hn}. 
The (black) line marked ``c" relaxes the assumption of KK mass degeneracy, and uses the actual MUED mass spectrum. 
Finally the dotted line is the result from the full calculation in MUED, including all co-annihilation processes, with the proper choice of masses. 
The green horizontal band denotes the preferred WMAP region for the relic density. 
We find good agreement between our calculation and existing results in literature, 
where the slight difference is due to the fact that existing results rely on non-relativistic velocity expansion, 
while we solve full Boltzmann equation \footnote{ We used MUED model from FeynRules web page:  http://feynrules.irmp.ucl.ac.be/wiki/MUED.}.
For the same reason, we are not able to reproduce the red curve (marked as ``a''), 
because it is impossible to undo the relativistic corrections in our code.

\section{Summary and conclusions}

\label{sec:conclusions}

We presented \verb|MadDM v.1.0|, a numerical code to calculate dark matter relic density in a generic \verb|MadGraph| model. The project is a pioneering effort to provide the \verb|MadGraph| framework with the ability to calculate cosmological and astro-physical signatures of dark matter.  The aim of the project is to bridge the existing gap between collider and dark matter numerical tools and thus provide the experimental collaborations with an ``all in one'' dark matter phenomenology package which can be easily incorporated into the future dark matter searches at the LHC.

We envisioned \verb|MadDM| as an easy-to-use, easy-to-install plugin for MadGraph.
\verb|MadDM| can be operated in a modular mode, whereby a user can use the existing Python libraries to perform a customized dark matter calculation or parameter scans. For convenience, we also provided a simple user interface which guides the user through the necessary steps to perform the relic density calculation.

Much like \verb|MadGraph 5| itself, \verb|MadDM| code is split into two modules. The Python module of the code utilizes the existing \verb|MadGraph 5| architecture to generate dark matter annihilation diagrams in any Universal FeynRules Output (UFO) model. A FORTRAN module then calculates the corresponding relic density. 

\verb|MadDM| takes into account the full effects of resonant-annihilation and co-annihilations. In addition, an important feature of \verb|MadDM| is that it is also able to calculate relic density in a general dark matter annihilation scenario, with any number of dark matter candidates.
We performed detailed validation of the code against \verb|micrOMEGAs| against several benchmark dark matter models. The real singlet extension of the Standard Model served to test the ability of \verb|MadDM| to handle resonant annihilation channels, while the complex singlet extension of the Standard Model offered a framework to test models with co-annihilations. In both cases we found excellent agreement between the codes. The exception is the region very close and above the resonance in the real singlet model where we noticed a discrepancy due to precision issues. The agreement between \verb|micrOMEGAs| and \verb|MadDM| can be improved by readjusting the precision parameters of the \verb|MadDM| numerical module. 

In addition to simplified models, we also tested \verb|MadDM| in the framework of MSSM and MUED. The existing literature on Kaluza-Klein dark matter agrees well with our results. The benchmark points of the MSSM show a good agreement with the results obtained with \verb|micrOMEGAs| in most cases, while a discrepancy of up to $30\%$ appears is parts of the parameter space where the details of the running $b$ mass produce significant effects. The large discrepancy is due to the differences in the treatment of running $b$ masses and the Higgs width in the spectrum generator used by \verb|micrOMEGAs| and the version of SUSPECT we used for comparison. 

In addition to relic density, future renditions of the \verb|MadDM| code will also provide the ability to compute direct detection and indirect detection rates, as well as relic density in models which allow for semi-annihilations.

\bigskip

\section*{Acknowledgmenents}
This work is supported in part by a US Department of Energy Grant Number DE-FG02-12ER41809,  
by the National Science Foundation under Award No. EPS-0903806 and 
by matching funds from the State of Kansas through Kansas Technology Enterprise Corporation. 
We are grateful to Olivier Mattelaer and Johan Alwall for helpful discussions and advice on the MadGraph 5 code and John Ralston for useful insight and suggestions.  

\appendix

\section{Example MadDM Python Script}

To illustrate the use of the \verb|MadDM| code, we proceed with an example. The following code illustrates a possible implementation of the \verb|MadDM| code which omits the user interface, yet uses most of the important features of the code. The program loads the model labeled by \verb|rsxSM-coann| (Two Real Singlet Extension of the Standard Model) and calculates the relic density. The dark matter candidate \verb|x1| and the co-annihilating particle \verb|x2| are hard-coded in this case. Removing these arguments would prompt the function to search for the dark matter / co-annihilation particles automatically. Finally, the code outputs the resulting relic density  on the screen:

\begin{verbatim} 
#! /usr/bin/env python
from init import *
from darkmatter import *

#Create the relic density object. 
dm=darkmatter()
#Initialize it from the rsxSM-coann model in the MadGraph model folder, 
#and store all the results in the Projects/rsxSM-coann subfolder. 
dm.init_from_model('rsxSM-coann', 'rsxSM-coann', new_proj = True)

# Determine the dark matter candidate...
dm.FindDMCandidate('x1')

#...and all the co-annihilation partners.
dm.FindCoannParticles('x2')

#Get the project name with the set of DM particles and see 
#if it already exists.
dm.GetProjectName()

#Generate all 2-2 diagrams.        
dm.GenerateDiagrams()    

#Print some dark matter properties in the mean time.
print "------ Testing the darkmatter object properties ------"
print "DM name: "+dm._dm_particles[0].get('name')
print "DM spin: "+str(dm._dm_particles[0].get('spin'))
print "DM mass var: "+dm._dm_particles[0].get('mass')
print "Mass: "+ str(dm.GetMass(dm._dm_particles[0].get('pdg_code')))
print "Project: "+dm._projectname

#Output the FORTRAN version of the matrix elements 
#and compile the numerical code.
dm.CreateNumericalSession()

#Calculate relic density.
omega = dm.CalculateRelicAbundance()
print "----------------------------------------------"
print "Relic Density: "+str(omega),
print "----------------------------------------------"

\end{verbatim}

\listoftables           
\listoffigures          


\bibliographystyle{model1-num-names}
\bibliography{manual}

\begin{thebibliography}{22}
\expandafter\ifx\csname natexlab\endcsname\relax\def\natexlab#1{#1}\fi
\providecommand{\url}[1]{\texttt{#1}}
\providecommand{\href}[2]{#2}
\providecommand{\path}[1]{#1}
\providecommand{\DOIprefix}{doi:}
\providecommand{\ArXivprefix}{arXiv:}
\providecommand{\URLprefix}{URL: }
\providecommand{\Pubmedprefix}{pmid:}
\providecommand{\doi}[1]{\href{http://dx.doi.org/#1}{\path{#1}}}
\providecommand{\Pubmed}[1]{\href{pmid:#1}{\path{#1}}}
\providecommand{\bibinfo}[2]{#2}
\ifx\xfnm\relax \def\xfnm[#1]{\unskip,\space#1}\fi
\bibitem[{Belyaev et~al.(2013)Belyaev, Christensen, and
  Pukhov}]{Belyaev:2012qa}
\bibinfo{author}{A.~Belyaev}, \bibinfo{author}{N.~D. Christensen},
  \bibinfo{author}{A.~Pukhov},
\newblock \bibinfo{title}{{CalcHEP 3.4 for collider physics within and beyond
  the Standard Model}},
\newblock \bibinfo{journal}{Comput.Phys.Commun.} \bibinfo{volume}{184}
  (\bibinfo{year}{2013}) \bibinfo{pages}{1729--1769}.
\bibitem[{Belanger et~al.(2007)Belanger, Boudjema, Pukhov, and
  Semenov}]{Belanger:2006is}
\bibinfo{author}{G.~Belanger}, \bibinfo{author}{F.~Boudjema},
  \bibinfo{author}{A.~Pukhov}, \bibinfo{author}{A.~Semenov},
\newblock \bibinfo{title}{{MicrOMEGAs 2.0: A Program to calculate the relic
  density of dark matter in a generic model}},
\newblock \bibinfo{journal}{Comput.Phys.Commun.} \bibinfo{volume}{176}
  (\bibinfo{year}{2007}) \bibinfo{pages}{367--382}.
\bibitem[{Alwall et~al.(2011)Alwall, Herquet, Maltoni, Mattelaer, and
  Stelzer}]{Alwall:2011uj}
\bibinfo{author}{J.~Alwall}, \bibinfo{author}{M.~Herquet},
  \bibinfo{author}{F.~Maltoni}, \bibinfo{author}{O.~Mattelaer},
  \bibinfo{author}{T.~Stelzer},
\newblock \bibinfo{title}{{MadGraph 5 : Going Beyond}},
\newblock \bibinfo{journal}{JHEP} \bibinfo{volume}{1106} (\bibinfo{year}{2011})
  \bibinfo{pages}{128}.
\bibitem[{Gleisberg et~al.(2009)Gleisberg, Hoeche, Krauss, Schonherr, Schumann
  et~al.}]{Gleisberg:2008ta}
\bibinfo{author}{T.~Gleisberg}, \bibinfo{author}{S.~Hoeche},
  \bibinfo{author}{F.~Krauss}, \bibinfo{author}{M.~Schonherr},
  \bibinfo{author}{S.~Schumann}, et~al.,
\newblock \bibinfo{title}{{Event generation with SHERPA 1.1}},
\newblock \bibinfo{journal}{JHEP} \bibinfo{volume}{0902} (\bibinfo{year}{2009})
  \bibinfo{pages}{007}.
\bibitem[{Gondolo et~al.(2004)Gondolo, Edsjo, Ullio, Bergstrom, Schelke
  et~al.}]{Gondolo:2004sc}
\bibinfo{author}{P.~Gondolo}, \bibinfo{author}{J.~Edsjo},
  \bibinfo{author}{P.~Ullio}, \bibinfo{author}{L.~Bergstrom},
  \bibinfo{author}{M.~Schelke}, et~al.,
\newblock \bibinfo{title}{{DarkSUSY: Computing supersymmetric dark matter
  properties numerically}},
\newblock \bibinfo{journal}{JCAP} \bibinfo{volume}{0407} (\bibinfo{year}{2004})
  \bibinfo{pages}{008}.
\bibitem[{Aoki et~al.(2012)Aoki, Duerr, Kubo, and Takano}]{Aoki:2012ub}
\bibinfo{author}{M.~Aoki}, \bibinfo{author}{M.~Duerr},
  \bibinfo{author}{J.~Kubo}, \bibinfo{author}{H.~Takano},
\newblock \bibinfo{title}{{Multi-Component Dark Matter Systems and Their
  Observation Prospects}},
\newblock \bibinfo{journal}{Phys.Rev.} \bibinfo{volume}{D86}
  (\bibinfo{year}{2012}) \bibinfo{pages}{076015}.
\bibitem[{Griest and Seckel(1991)}]{Griest:1990kh}
\bibinfo{author}{K.~Griest}, \bibinfo{author}{D.~Seckel},
\newblock \bibinfo{title}{{Three exceptions in the calculation of relic
  abundances}},
\newblock \bibinfo{journal}{Phys.Rev.} \bibinfo{volume}{D43}
  (\bibinfo{year}{1991}) \bibinfo{pages}{3191--3203}.
\bibitem[{Ibe et~al.(2009)Ibe, Murayama, and Yanagida}]{Ibe:2008ye}
\bibinfo{author}{M.~Ibe}, \bibinfo{author}{H.~Murayama},
  \bibinfo{author}{T.~Yanagida},
\newblock \bibinfo{title}{{Breit-Wigner Enhancement of Dark Matter
  Annihilation}},
\newblock \bibinfo{journal}{Phys.Rev.} \bibinfo{volume}{D79}
  (\bibinfo{year}{2009}) \bibinfo{pages}{095009}.
\bibitem[{Guo and Wu(2009)}]{Guo:2009aj}
\bibinfo{author}{W.-L. Guo}, \bibinfo{author}{Y.-L. Wu},
\newblock \bibinfo{title}{{Enhancement of Dark Matter Annihilation via
  Breit-Wigner Resonance}},
\newblock \bibinfo{journal}{Phys.Rev.} \bibinfo{volume}{D79}
  (\bibinfo{year}{2009}) \bibinfo{pages}{055012}.
\bibitem[{Srednicki et~al.(1988)Srednicki, Watkins, and
  Olive}]{Srednicki:1988ce}
\bibinfo{author}{M.~Srednicki}, \bibinfo{author}{R.~Watkins},
  \bibinfo{author}{K.~A. Olive},
\newblock \bibinfo{title}{{Calculations of Relic Densities in the Early
  Universe}},
\newblock \bibinfo{journal}{Nucl.Phys.} \bibinfo{volume}{B310}
  (\bibinfo{year}{1988}) \bibinfo{pages}{693}.
\bibitem[{Djouadi et~al.(2007{\natexlab{a}})Djouadi, Kneur, and
  Moultaka}]{Djouadi:2002ze}
\bibinfo{author}{A.~Djouadi}, \bibinfo{author}{J.-L. Kneur},
  \bibinfo{author}{G.~Moultaka},
\newblock \bibinfo{title}{{SuSpect: A Fortran code for the supersymmetric and
  Higgs particle spectrum in the MSSM}},
\newblock \bibinfo{journal}{Comput.Phys.Commun.} \bibinfo{volume}{176}
  (\bibinfo{year}{2007}{\natexlab{a}}) \bibinfo{pages}{426--455}.
\bibitem[{Djouadi et~al.(2007{\natexlab{b}})Djouadi, Muhlleitner, and
  Spira}]{Djouadi:2006bz}
\bibinfo{author}{A.~Djouadi}, \bibinfo{author}{M.~Muhlleitner},
  \bibinfo{author}{M.~Spira},
\newblock \bibinfo{title}{{Decays of supersymmetric particles: The Program
  SUSY-HIT (SUspect-SdecaY-Hdecay-InTerface)}},
\newblock \bibinfo{journal}{Acta Phys.Polon.} \bibinfo{volume}{B38}
  (\bibinfo{year}{2007}{\natexlab{b}}) \bibinfo{pages}{635--644}.
\bibitem[{Belanger et~al.(2011)Belanger, Boudjema, Brun, Pukhov, Rosier-Lees
  et~al.}]{Belanger:2010gh}
\bibinfo{author}{G.~Belanger}, \bibinfo{author}{F.~Boudjema},
  \bibinfo{author}{P.~Brun}, \bibinfo{author}{A.~Pukhov},
  \bibinfo{author}{S.~Rosier-Lees}, et~al.,
\newblock \bibinfo{title}{{Indirect search for dark matter with
  micrOMEGAs2.4}},
\newblock \bibinfo{journal}{Comput.Phys.Commun.} \bibinfo{volume}{182}
  (\bibinfo{year}{2011}) \bibinfo{pages}{842--856}.
\bibitem[{Barger et~al.(2008)Barger, Langacker, McCaskey, Ramsey-Musolf, and
  Shaughnessy}]{Barger:2007im}
\bibinfo{author}{V.~Barger}, \bibinfo{author}{P.~Langacker},
  \bibinfo{author}{M.~McCaskey}, \bibinfo{author}{M.~J. Ramsey-Musolf},
  \bibinfo{author}{G.~Shaughnessy},
\newblock \bibinfo{title}{{LHC Phenomenology of an Extended Standard Model with
  a Real Scalar Singlet}},
\newblock \bibinfo{journal}{Phys.Rev.} \bibinfo{volume}{D77}
  (\bibinfo{year}{2008}) \bibinfo{pages}{035005}.
\bibitem[{Barger et~al.(2009)Barger, Langacker, McCaskey, Ramsey-Musolf, and
  Shaughnessy}]{Barger:2008jx}
\bibinfo{author}{V.~Barger}, \bibinfo{author}{P.~Langacker},
  \bibinfo{author}{M.~McCaskey}, \bibinfo{author}{M.~Ramsey-Musolf},
  \bibinfo{author}{G.~Shaughnessy},
\newblock \bibinfo{title}{{Complex Singlet Extension of the Standard Model}},
\newblock \bibinfo{journal}{Phys.Rev.} \bibinfo{volume}{D79}
  (\bibinfo{year}{2009}) \bibinfo{pages}{015018}.
\bibitem[{Barger et~al.(2010)Barger, McCaskey, and Shaughnessy}]{Barger:2010yn}
\bibinfo{author}{V.~Barger}, \bibinfo{author}{M.~McCaskey},
  \bibinfo{author}{G.~Shaughnessy},
\newblock \bibinfo{title}{{Complex Scalar Dark Matter vis-\`{a}-vis CoGeNT,
  DAMA/LIBRA and XENON100}},
\newblock \bibinfo{journal}{Phys.Rev.} \bibinfo{volume}{D82}
  (\bibinfo{year}{2010}) \bibinfo{pages}{035019}.
\bibitem[{Allanach et~al.(2002)Allanach, Battaglia, Blair, Carena, De~Roeck
  et~al.}]{Allanach:2002nj}
\bibinfo{author}{B.~Allanach}, \bibinfo{author}{M.~Battaglia},
  \bibinfo{author}{G.~Blair}, \bibinfo{author}{M.~S. Carena},
  \bibinfo{author}{A.~De~Roeck}, et~al.,
\newblock \bibinfo{title}{{The Snowmass points and slopes: Benchmarks for SUSY
  searches}},
\newblock \bibinfo{journal}{Eur.Phys.J.} \bibinfo{volume}{C25}
  (\bibinfo{year}{2002}) \bibinfo{pages}{113--123}.
\bibitem[{Degrande et~al.(2012)Degrande, Duhr, Fuks, Grellscheid, Mattelaer
  et~al.}]{Degrande:2011ua}
\bibinfo{author}{C.~Degrande}, \bibinfo{author}{C.~Duhr},
  \bibinfo{author}{B.~Fuks}, \bibinfo{author}{D.~Grellscheid},
  \bibinfo{author}{O.~Mattelaer}, et~al.,
\newblock \bibinfo{title}{{UFO - The Universal FeynRules Output}},
\newblock \bibinfo{journal}{Comput.Phys.Commun.} \bibinfo{volume}{183}
  (\bibinfo{year}{2012}) \bibinfo{pages}{1201--1214}.
\bibitem[{Servant and Tait(2002)}]{Servant:2002hb}
\bibinfo{author}{G.~Servant}, \bibinfo{author}{T.~M. Tait},
\newblock \bibinfo{title}{{Elastic scattering and direct detection of
  Kaluza-Klein dark matter}},
\newblock \bibinfo{journal}{New J.Phys.} \bibinfo{volume}{4}
  (\bibinfo{year}{2002}) \bibinfo{pages}{99}.
\bibitem[{Arrenberg et~al.(2008)Arrenberg, Baudis, Kong, Matchev, and
  Yoo}]{Arrenberg:2008wy}
\bibinfo{author}{S.~Arrenberg}, \bibinfo{author}{L.~Baudis},
  \bibinfo{author}{K.~Kong}, \bibinfo{author}{K.~T. Matchev},
  \bibinfo{author}{J.~Yoo},
\newblock \bibinfo{title}{{Kaluza-Klein Dark Matter: Direct Detection vis-a-vis
  LHC}},
\newblock \bibinfo{journal}{Phys.Rev.} \bibinfo{volume}{D78}
  (\bibinfo{year}{2008}) \bibinfo{pages}{056002}.
\bibitem[{Kong and Matchev(2006)}]{Kong:2005hn}
\bibinfo{author}{K.~Kong}, \bibinfo{author}{K.~T. Matchev},
\newblock \bibinfo{title}{{Precise calculation of the relic density of
  Kaluza-Klein dark matter in universal extra dimensions}},
\newblock \bibinfo{journal}{JHEP} \bibinfo{volume}{0601} (\bibinfo{year}{2006})
  \bibinfo{pages}{038}.
\bibitem[{Arrenberg et~al.(2013)Arrenberg, Baudis, Kong, Matchev, and
  Yoo}]{Arrenberg:2013paa}
\bibinfo{author}{S.~Arrenberg}, \bibinfo{author}{L.~Baudis},
  \bibinfo{author}{K.~Kong}, \bibinfo{author}{K.~T. Matchev},
  \bibinfo{author}{J.~Yoo},
\newblock \bibinfo{title}{{Kaluza-Klein Dark Matter: Direct Detection vis-a-vis
  LHC (2013 update)}}  (\bibinfo{year}{2013}).

\end{thebibliography}


\end{document}